\documentclass[useAMS,usenatbib]{mn2e}
\usepackage{subfigure}
\usepackage{natbib}
\usepackage{lscape}
\usepackage[dvips]{color}
\usepackage{amssymb}
\usepackage{graphicx}
\usepackage{ulem} 
\usepackage{url} 

\title[The Evolution of ${\it K}^{\ast}$ and the HOD since $z=1.5$]{The Evolution of ${\it K}^{\ast}$ and the Halo Occupation Distribution since $z=1.5$: Observations vs. Simulations}
\author[Capozzi D. et al. 2011]
 { Diego Capozzi $^{1}$\thanks{E-mail:dc@astro.livjm.ac.uk}, Chris A. Collins $^{1}$, John P. Stott $^{1,2}$ and Matt Hilton$^{1,3}$ \\ \\
 1 - Astrophysics Research Institute, Liverpool John Moores University, Twelve Quays House, Egerton Wharf,\\
  \quad \, Birkenhead, CH41 1LD\\
  2 - Extragalactic \& Cosmology Group, Department of Physics, University of Durham, South Road, Durham DH1 3LE\\
  3 - School of Physics and Astronomy, University of Nottingham, NG7 2RD\\
}
\date{Accepted ;
      Received ;
      in original form }
\pagerange{\pageref{firstpage}--\pageref{lastpage}} \pubyear{2011}

\begin{document}

\maketitle

\label{firstpage}

\begin{abstract}
We study the evolution of the {\it K}-band luminosity function (LF) and the halo occupation 
distribution (HOD) using Subaru observations of 15 X-ray clusters at $z=0.8-1.5$ and compare the results with mock clusters ($0<z<1.3$) extracted from the Millennium Simulation
and populated with galaxies by means of the semi-analytic model (SAM) of Bower et al.,  matched in mass to our observed sample. 
By fixing the faint-end slope ($\alpha=-0.9$), we find that the characteristic luminosity 
${\it K}^{\ast}$ defined by a Shechter LF is consistent with the predictions of the SAM, which are found, for the first time, to mimic well the evolution of ${\it K}^{\ast}$ in rich clusters at $z\geq 1$. However, we cannot distinguish between this model and a simple stellar population synthesis model invoking passive evolution with a formation redshift ($z_{\rm f}\simeq5$) - consistent with the presence of an old red galaxy population ubiquitous in rich clusters at $z=1.5$.
We also see a small difference ($\Delta {\it K}^{\ast} \simeq 0.5$) between our clusters and studies of the field population at similar redshifts, which suggests only a weak dependence of the luminous ($L\geq L^{\ast}$) part of the LF
on cluster environment. Turning to our study of the HOD, we find that within a radius corresponding to a density 500 times critical, high-$z$ clusters tend to host smaller numbers of galaxies to a magnitude ${\it K}^{\ast}+2$ compared to their low-$z$ counterparts.  This behavior is also seen in the mock samples and is relatively insensitive to the average mass of the cluster haloes.
In particular, we find significant correlations of the observed number of member cluster galaxies ($N$) with both $z$ and cluster mass:
$N(M,z)=(53\pm1)(1+z)^{-0.61^{+0.18}_{-0.20}} (M/10^{14.3})^{0.86\pm0.05}$.
Finally, we examine the spatial distribution of galaxies and provide a new estimate of the concentration parameter for clusters at high redshift ($c_{\rm g}=2.8^{+1.0}_{-0.8}$). 
Our result is consistent with predictions from both our SAM mock clusters and predictions of dark matter haloes from the literature. The mock sample predictions rise slowly with decreasing redshift reaching $c_{\rm g}=6.3^{+0.39}_{-0.36}$ at $z=0$. 
\end{abstract}


\section{INTRODUCTION}
\label{sec:section1}
The cold dark matter (CDM) paradigm is able to predict the formation of structures, which form through gravitational instabilities and
cluster hierarchically. A way to study the structure formation, is by using the halo model formalism, according 
to which all galaxies in the Universe dwell in virialised units of mass called haloes which obey universal scaling relations. 
 N-body simulations \citep{Navarro-2004} show that dark matter haloes seem 
to have universal velocity and density profiles. The same cannot be said for galaxies, which are observed to be biased 
tracers of the mass distribution and whose clustering amplitude seems to depend on their properties. Studying the bias can provide insight into the physics of galaxy formation and the relationship between the galaxy and dark matter distributions, with the aim of uncovering the processes responsible for the way in which haloes, whose properties are specified by the cosmological model, are populated with galaxies \citep{Berlind-2002}. \\

A fundamental tool for investigating the influence of bias on galaxy clustering statistics is the halo occupation 
distribution (HOD), generally used to determine the average number of galaxies within a dark matter halo as a function of the halo mass. 
Within the HOD framework, the virialised dark matter haloes with typical overdensities of $\Delta \sim 200$ (defined with respect to the critical density) are
expected to be in approximate dynamical equilibrium and are described in one of three ways: (i) P(N$|$M), the probability of a halo of given mass $M$ having an average 
number of galaxies $\langle N\rangle$; (ii) the relation between galaxy and dark matter spatial distributions within haloes and (iii) the relation
between galaxy and dark matter velocity distributions within haloes. Theoretical studies (e.g.,  \citealp{Peacock-2000, Benson-2000a, Berlind-2002, Berlind-2003, Kravtsov-2004}) 
based on N-body simulations, hydrodynamic simulations and semi-analytic models (SAMs) showed that P(N$|$M) can be well modeled by a Poissonian 
distribution in the high halo mass regime ($M_{\rm h}\gtrsim 10^{12}\ M_{\odot}$) while it assumes significant sub-Poissonian behavior, which can 
be modeled by a ``nearest integer \footnote{Referred to as ``Average'' in \citet{Benson-2000a} and \citet{Berlind-2002}. The 
definition of this distribution is $p(N_{\rm l}|\langle N\rangle)=1-(\langle N\rangle-N_{\rm l})$, $p(N_{\rm l}+1|\langle N\rangle)=\langle N\rangle-N_{\rm l}$, 
where $N_{\rm l}$ is the integer satisfying $N_{\rm l}\leq\langle N\rangle<N_{\rm l}+1$, with $p(N|\langle N\rangle)$=0 for all other values of N.} (Nint) 
distribution instead, for lower halo masses. These studies also showed $\langle N\rangle_{\rm M}$ presents a sharp cutoff at low halo mass, a slowly 
rising plateau and a steep increase of the occupation number at high halo mass. A simple way to model the complicated shape of 
$\langle N\rangle_{\rm M}$, avoiding the use of models involving a large number of parameters, is by assuming 
the existence of two separate galaxy populations within haloes: (i) central galaxies and (ii) satellite galaxies. This choice is motivated by
reasons based on hydrodynamic simulations \citep{Berlind-2003} and on studies of observed galaxy clusters and groups, which take the brightest cluster galaxies (BCGs)
as a different population from the rest of the cluster galaxies. 
These two populations can be modeled separately (e.g.,  \citealp{Benson-2000a, Kravtsov-2004, Zheng-Coil-2007}), with the simplest case being modeling the HOD of central galaxies as 
a step function $\langle N_{\rm c}\rangle=1$ above a halo mass limit and a power law for satellite galaxies. In this way the HOD becomes
a measure of the combined probability that a halo of mass $M$ hosts a central galaxy and a given number $N_{\rm s}$ of 
satellite galaxies.\\ 
Studies have also been carried out on the radial, spatial and velocity distributions of galaxies and dark matter, the majority of them focused
on the latter. For instance, the halo concentration (defined as one of the shape parameters of the radial density profile) and its relation with
mass and redshift have been the topic of many investigations in recent years (e.g. \citealp{Nagai-2005, Neto-2007, Gao-2008, Munoz-2011}). 
The importance of the concentration resides
in its connection with the mean density of the Universe at the time of collapse, i.e. more concentrated structures formed earlier in the past, when the Universe was denser (\citealp{Navarro-1997} hereafter NFW, \citealp{Neto-2007, Gao-2008, Munoz-2011}). 
For this reason, understanding the dependence of concentration on mass and redshift and how the relationship between the concentration of dark matter 
and galaxies within haloes evolves with cosmic time, can provide fundamental insight into the formation and evolution of structures and into galaxy formation
process.\\

Several statistical galaxy properties have been used to empirically measure the HOD assuming a power-law form (${\langle N \rangle}_{\rm M} \propto M^{\beta}$) for satellite galaxies. For instance, 
studies of the HOD have been carried out by using either the luminosity function (e.g., \citealp{Yang-2008}), the spatial clustering (e.g., \citealp{Phleps-2006,Abbas-2010}) or 
by counting galaxies within known dark matter haloes, such as rich clusters (e.g., \citealp{Kochanek-2003, Lin-2004, Collister-Lahav-2005, Popesso-2007, Ho-2009}). Our study focuses on this latter method, which is perhaps the most direct,  using a sample of 15 X-ray selected clusters at high redshift. \\

Unfortunately, there is a significant disagreement among the results so far obtained and firm conclusions are hard to draw. For example,  
the slope of the $N-M$ relation is seen ranging between $\beta=0.55$ \citep{Marinoni-2002} 
and $\beta \sim 1.7$ \citep{Abbas-2010}; while the average concentration parameter of the galaxy density profile ($c_{\rm g}$) is found to range between  
$c_{\rm g}\sim 2$ and $c_{\rm g}\sim 8$ \citep{Carlberg-1997, van-der-Marel-2000, Biviano-2003, Katgert-2004, Lin-2004, Rines-2006, Muzzin-2007a, Biviano-2010}. Although the majority of studies have been 
conducted at low redshift, these problems of consistency among different studies of the HOD extend to intermediate redshifts (e.g., \citealp{Lin-2006, Buote-2007, Abbas-2010}), therefore 
there is a continuing motivation to investigate the evolution of the HOD with new galaxy samples.\\

The {\it K}-band LF can be used as a surrogate of the galaxy 
mass function because it is a sensitive probe of the bulk properties of galaxy populations out to $z\simeq 1.5$. There are several advantages to using  
the near-infrared light for such studies: i) {\it K}-band ($2.2\ {\rm \mu m} $) luminosities broadly reflect the total stellar mass of galaxies, resulting in a M/L ratio that is insensitive 
to the star-formation history of early-type galaxies; ii) stars are easy to remove, 
as they generally have ${\it J}-{\it K}<1$ (Vega), whilst the k-correction makes the observed colours of the great majority of galaxies ${\it J}-{\it K}>1$ 
\citep{De-Propris-1999, McCracken-2010}; iii) k-correction in the near-infrared bands varies slowly with $z$ and depends weakly on Hubble type 
\citep{Poggianti-1997, Bruzual-2003}; iv) the effect of extinction at these wavelengths is significantly smaller than in optical and UV passbands.
With the advent of infrared surveys like the {\it UKIRT Infrared Deep Sky Survey} (UKIDSS, \citealp{Lawrence-2007}) and Spitzer, it has been possible to carry out studies of the stellar mass of high-$z$ 
galaxies. However, such studies produced results in contrast with the prediction of SAMs. For instance, \citet{De-Propris-1999, Ellis-2004, Strazzullo-2006} and \citet{Lin-2006} studied the 
evolution of the cut-off magnitude ($K^{\ast}$) of the cluster {\it K}-band LF out to $z\sim 1$ and found agreement with passive evolving models which have 
formation redshift ($z_{\rm f} \sim 2-5$), suggesting that the bulk stellar mass of ${\it K}^{\ast}$ cluster galaxies has not increased 
substantially since $z=1$. Several other studies showed that the high-mass end of the galaxy mass function seems to remain pretty much unchanged since $z\sim1$ for elliptical galaxies (e.g., \citealp{Cimatti-2006} and \citealp{Pozzetti-2010}) and similar results have been obtained for BCGs (\citealp{Whiley-2008,Collins-2009} and \citealp{Stott-Collins-2010}). These results suggest a timescale for the mass assemblage of galaxies similar to the age of their component 
stars, consistent with a monolithic-like model, and such activity can be viewed at least as qualitatively as consistent with a ``downsizing'' (\citealp{Cowie-1996, Thomas-2005, De-Lucia-2007, Stott-2007, Capozzi-2010} and references therein) process, according to which the more massive early-type galaxies
end their star formation and settle on the colour-magnitude relation earlier than their less massive counterparts.

By contrast, the  SAM used by \citet{De-Lucia-2006} predicts that the majority ($\simeq 70$ per cent) of 
the stellar mass of ellipticals at $z=0$ is already formed by $z=1$, but that at this redshift only a few per cent of this mass is assembled in the main progenitor.\\ 

In this paper we investigate the process of galaxy formation by 
studying the {\it K}-band LF (evolution of $K^{\ast}$) and the HOD of a sample of 15 galaxy clusters, containing the majority of the highest-$z$ ($z>0.8$)
X-ray clusters observed so far. We want to push the study of the evolution of cluster galaxies to higher $z$ by extending the Hubble diagram of 
${\it K}^{\ast}$ out to $z\sim 1.5$, because it is at $z\gtrsim 0.8$ that the differences among the evolutionary predictions based on stellar populations 
models are enhanced. The HOD of this sample is investigated using the number ($N_{500}$) of 
cluster galaxies within $R_{500}$ (the radial distance where $\Delta=500$) as a function of cluster mass and $z$. 
We also study galaxy clustering by analysing 
the radial galaxy surface number density profile in order to estimate the galaxy concentration parameter at $z\sim 1$ to compare with similar estimates at low $z$ (e.g., \citealp{Lin-2004}).
We compare the results of our HOD analysis with results obtained from dark matter haloes of similar redshift selected from the Millennium Simulation 
(MS; \citealp{Springel-2005}), whose haloes are populated with galaxies taken from the SAM by \citet{Bower-2006}. It is worth pointing out that the majority 
of previous observation-based studies estimated $c_{\rm g}$ at low $z$ (e.g, \citealp{Lin-2004, Popesso-2007}), apart from some recent exceptions \citep{Biviano-2010}. In addition, 
even though more attention to the concentration parameter has been given by theoretical studies based on N-body simulations (e.g. \citealp{Neto-2007, Gao-2008, Duffy-2008}) and 
SAM \citep{Nagai-2005}, the majority of them focused more on $c_{\rm dm}$ (the concentration parameter of dark matter haloes' radial density profiles) and its relation with halo mass and $z$. Here, instead, simulations and their associated SAMs are used to study the concentration of galaxies $c_{\rm g}$. To our knowledge this is the first attempt to carry out a self-consistent
comparison between the observed galaxy concentration parameter in clusters
and that in mock clusters within the MS.\\

The paper is organised as follows: in Section \ref{sec:section2} we describe the observed cluster sample, while Section \ref{sec:section3} is 
dedicated to data reduction and photometry. Sections \ref{sec:section4} and \ref{sec:section5} are focused on the study of the {\it K}-band LF and 
the HOD respectively of the observed and mock samples, while in Sections \ref{sec:section6} and \ref{sec:section7} we present and discuss 
our results. Finally, we draw our conclusions in Section \ref{sec:section8}.\\  
        
Throughout this paper we make use of magnitudes in the Vega photometric system and assume a standard cosmology with 
$H_{0}=70\ {\rm km\ s^{-1}\ Mpc^{-1}}$, $\Omega_{\rm m}=0.3$ and $\Omega_{\rm \Lambda}=0.7$.  

\section{Cluster Sample}
\label{sec:section2}
\begin{figure}
\begin{center}
\includegraphics[width=0.5\textwidth,angle=0,clip]{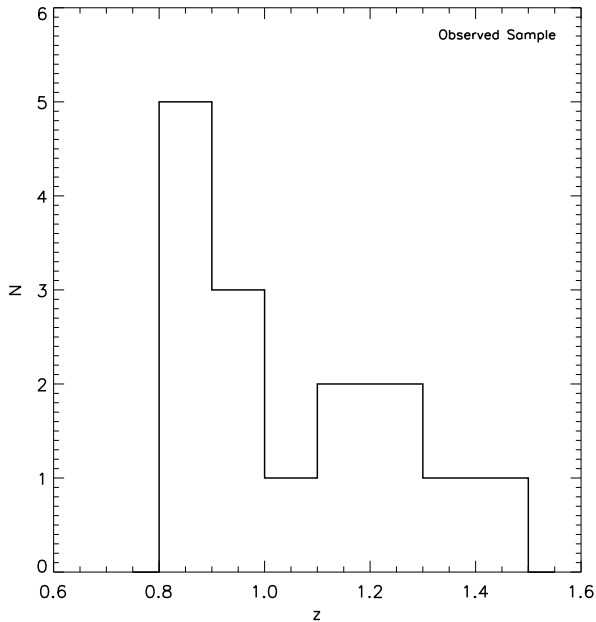} 
\caption{Spectroscopic redshift distribution of the 15 observed clusters in our sample.}
\label{fig:Figure1}
\end{center}
\end{figure}

The clusters are a subsample of the cluster sample described in \citet{Stott-Collins-2010} and consist of the clusters from that list with available $J$ and $K$ band data. 
Two clusters (CL J1226+3332, \citealp{Maughan-2004}, and MS1054.4-0321, \citealp{Branchesi-2007}) were included in the 
original sample, but we decided to exclude them because the {\it MOIRCS} field of view (FOV) ($4' \times 7'$) (see below) was not extended enough to contain 
their large $R_{500}$. Our final sample consists of the 15 clusters between $0.8<z<1.5$  detailed in Table \ref{tab:Table1}. Some of these were discovered 
by various X-ray surveys whilst some were optically selected clusters showing extended X-ray emission. All 15 
clusters have spectroscopically confirmed redshifts (Fig. \ref{fig:Figure1}) and X-ray luminosities in the range 
$1<L_{\rm{X}}<19 \times 10^{44}\ {\rm {erg\ s^{-1}}}$.  Cluster mass estimates are made using the 
$M-T_{\rm {X}}$ relation in \citet{Stott-Collins-2010}, whose parameter values are based on the \citet{Maughan-2007} derived 
$M-T_{\rm {X}}$ relation. We refer to the study by \citet{Stott-Collins-2010} for further details about the derivation of cluster masses. 

\begin{figure}
\begin{center}
\includegraphics[width=0.5\textwidth,angle=0,clip]{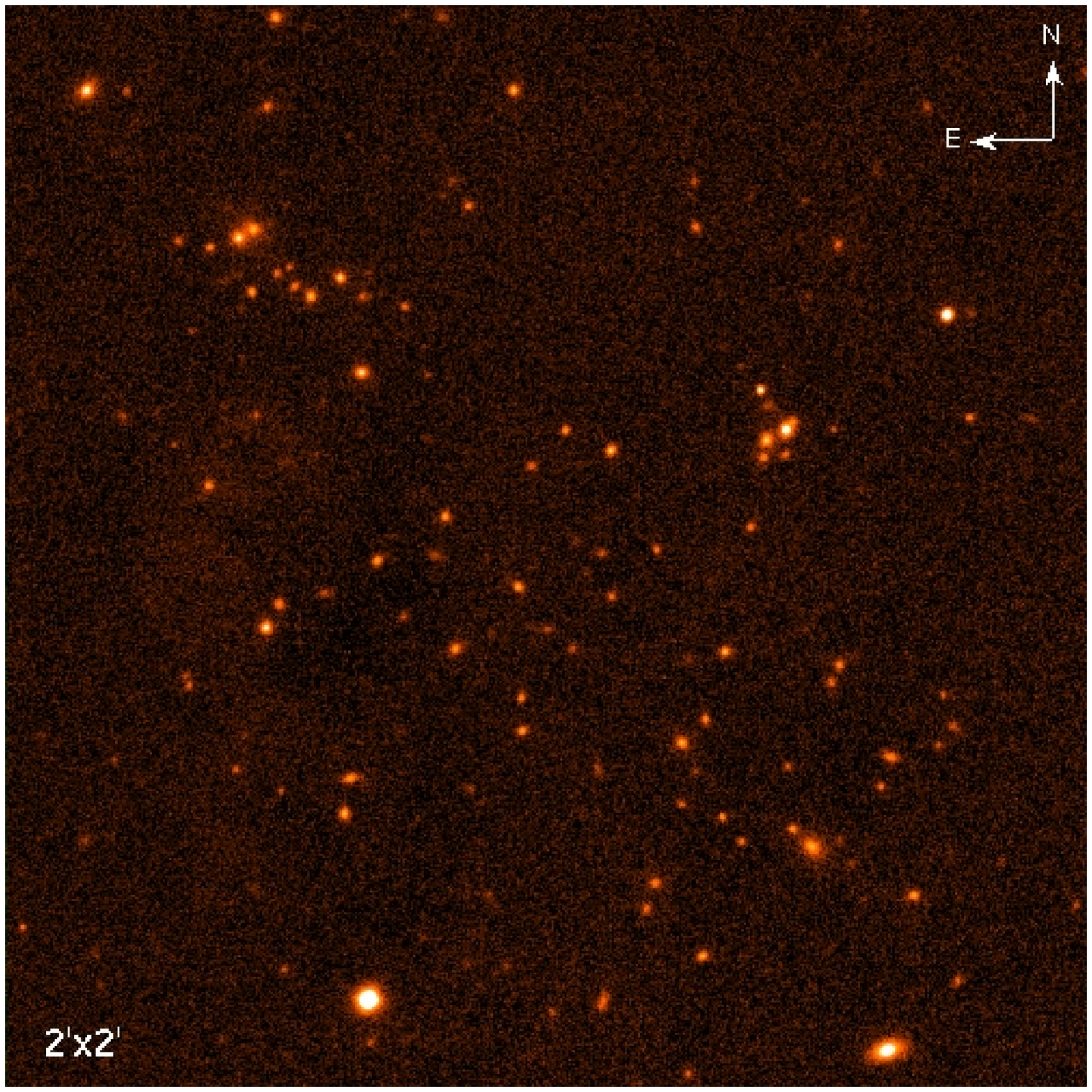} 
\caption{$2'\times 2'$ {\it K}-band image of CL J0152.7-1357}
\label{fig:Figure2}
\end{center}
\end{figure}

\section{Observations, Data Reduction and Photometry}
\label{sec:section3}
The observations were taken with the {\it MOIRCS} camera \citep{Ichikawa-2006} on the 8.2 {\rm m} {\it Subaru} telescope, which provides imaging and 
low-resolution spectroscopy over a total FOV of $4'\times 7'$ with a pixels scale of $0.117''$ per pixel. 
Observations were taken in $0.5''$ seeing on the nights of August 8th and 9th 2007 and in $0.3''-0.6''$
seeing on the nights of December 16th 2008 and April 18th 2009, with the clusters centered on Detector 2. A circular 11-point dither 
pattern of radius $25''$ was used for both bands to ensure good sky-subtraction. The modal integration times were 25 min at {\it J} and
21 min at ${\it K_{\rm {s}}}$, although we observed some of the higher redshift clusters for 50 per cent longer when scheduling allowed.
These exposures reach a $5\sigma$ limiting magnitude of at least {\it J}=21 and ${\it K_{\rm {s}}}=20$ (Vega).

The data are reduced using the external IRAF package MCSRED. They are flat-fielded, sky subtracted, corrected for distortion caused
by the camera optical design and registered to a common pixel coordinate system. The final reduced images on which we perform the 
photometry are made by taking the $3\sigma$ (s.d.) clipped mean of the dither frames. The galaxy photometry is extracted using 
SExtractor (version 2.5) MAG\_AUTO magnitude, which is found to be within $\sim 0.1$ mag of the total for extended sources 
\citep{Martini-2001}. To calculate the colours of the galaxies, we run SExtractor in dual image mode so that the ${\it K_{\rm {s}}}$ 
(hereafter {\it K}, for simplicity) band detections extract the {\it J} band catalogue with identical positions and apertures to ensure 
accurate colour determination. This photometry is carried out by using SExtractor MAG\_APER magnitude using a circular aperture of diameter $1''$.\\

The photometry is calibrated to the Vega system using a combination of standard star observations and the {\it Two Micron All Sky Survey} 
(2MASS, \citealp{Skrutskie-2006}) and UKIDSS catalogues. The typical photometric errors are 0.01 and 0.08 for the standard star and 
survey calibrated data, 

\begin{table*}
\begin{center}
\begin{tabular}{lllcccl}
\hline  
{\bf Name}  				  & {\bf RA}		&  {\bf Dec}	 &  {\bf z} &   $\mathbf{T_{\rm x}\ ({\rm {\bf keV}})}$	&  	$\mathbf{M_{200}\ (10^{14}\ {\rm {\bf M_{\mathbf{\odot}}}})}$	&  {\bf Reference}			   \\  					
\hline  
RDCS J1317 + 2911    	  & 13:17:21.70	&  +29:11:18.0  &  0.81 	&   $4.0_{-0.8}^{+1.3}$ 						&  	$2.7_{-1.3}^{+2.9}$  										 			&  \citet{Branchesi-2007}		\\
CL J0152.7 - 1357	   	  & 01:52:41.00	&  -13:57:45.0  &  0.83 	&   $5.4_{-0.9}^{+0.9}$ 						&  	$4.5_{-2.2}^{+2.7}$  										 			&  \citet{Vikhlinin-2009}		\\
CL J1559.1 + 6353   		  & 15:59:06.00	&  +63:53:00.0  &  0.85 	&   $4.1_{-1.0}^{+1.4}$ 						&  	$2.8_{-1.5}^{+3.2}$  										 			&  \citet{Maughan-2006}			\\
CL J1008.7 + 5342	   	  & 10:08:42.00	&  +53:42:00.0  &  0.87 	&   $3.6_{-0.6}^{+0.8}$ 						&  	$2.2_{-1.0}^{+1.6}$  										 			&  \citet{Maughan-2006}			\\
CL 1604 + 4304	   		  & 16:04:25.20	&  +43:04:53.0  &  0.9  	&   $2.5_{-0.7}^{+1.1}$ 						&  	$1.2_{-0.6}^{+1.6}$  										 			&  \citet{Lubin-2004} 			\\
CL J1429.0 + 4241     	  & 14:29:06.40	&  +42:41:10.0  &  0.92 	&   $6.2_{-1.0}^{+1.5}$ 						&  	$5.5_{-2.0}^{+5.3}$  										 			&  \citet{Maughan-2006} 		\\
RCS J0439 - 2904    		  & 04:39:38.00	&  -29:04:55.0  &  0.95 	&   $1.5_{-0.2}^{+0.3}$ 						&  	$0.5_{-0.2}^{+0.4}$  										 			&  \citet{Hicks-2008} 			\\
2XMM J083026 + 524133     & 08:30:25.90	&  +52:41:33.0  &  0.99 	&   $8.2_{-0.9}^{+0.9}$ 						&  	$8.5_{-3.4}^{+4.1}$  										 			&  \citet{Lamer-2008} 			\\
WARPS J1415.1 + 3612   	  & 14:15:11.10	&  +36:12:03.0  &  1.03 	&   $6.2_{-0.7}^{+0.8}$ 						&  	$5.2_{-1.9}^{+2.9}$  										 			&  \citet{Branchesi-2007} 		\\
RDCS J0910 + 5422   		  & 09:10:44.90	&  +54:22:09.0  &  1.11 	&   $6.4_{-1.2}^{+1.5}$ 						&  	$5.3_{-2.5}^{+4.1}$  										 			&  \citet{Balestra-2007} 		\\
RX J1053.7 + 5735 (West)  & 10:53:39.80	&  +57:35:18.0  &  1.14 	&   $4.4_{-0.3}^{+0.3}$ 						&  	$2.7_{-1.0}^{+1.4}$  										 			&  \citet{Hashimoto-2004} 		\\
XLSS J022303.0 - 043622   & 02:23:03.00	&  -04:36:22.0  &  1.22 	&   $3.5_{-0.4}^{+0.4}$ 						&  	$1.8_{-0.7}^{+0.9}$  										 			&  \citet{Stott-Collins-2010}, \citet{Bremer-2006} 			\\
RDCS J1252.9 - 2927    	  & 12:52:54.40	&  -29:27:17.0  &  1.24 	&   $7.2_{-0.6}^{+0.4}$ 						&  	$6.1_{-2.4}^{+2.3}$  										 			&  \citet{Balestra-2007} 		\\
XMMU J2235.3 - 2557  	  & 22:35:20.60	&  -25:57:42.0  &  1.39 	&   $8.6_{-1.2}^{+1.3}$ 						&  	$7.7_{-3.1}^{+4.4}$  										 			&  \citet{Rosati-2009} 			\\
XMMXCS J2215.9 - 1738  	  & 22:15:58.50	&  -17:38:03.0  &  1.46 	&   $4.1_{-0.9}^{+0.6}$ 						&  	$2.1_{-0.8}^{+1.9}$  										 			&  \citet{Hilton-2010} 		\\
\hline																																									\end{tabular}
\caption{The observed cluster sample.}
\label{tab:Table1}							    
\end{center}
\end{table*}

\noindent respectively.

In Figure \ref{fig:Figure2} the image of CL J0152.7-1357 obtained after all the described process is shown as an example.

\section{Analysis of the Observed Cluster Sample}
\label{sec:section4}
In order to carry out our study of the LF and the HOD, both foreground and background objects have to be removed or accounted for. It is well known 
(e.g., \citealp{Leggett-1992, De-Propris-1999}) that stars generally have observed ${\it {J-K}}<1$ (Vega), while, because of k-correction, the great majority 
of galaxies, excluding the most local ones, lie in the region ${\it {J-K}}>1$. Carrying out such a colour cut effectively removes contamination by stars. However, there is the possibility that faint blue galaxies have observed ${\it {J-K}}<1$, so for this reason 
we only perform this ${\it {J-K}}<1$ at magnitudes brighter than ${\it K}=18$, i.e. at least 2 mag brighter than the 5$\sigma$ limit of 
each cluster. Fainter than this, field galaxy counts outnumber those of the stars by more than a factor of 10 \citep{De-Propris-1999,McCracken-2010}, thus at these faint 
magnitudes the correction for stellar contamination becomes relatively unimportant. From simulations we estimate a contamination of only $3-4$ stars per cluster at ${\it K}<18$, which remains stable down to a 4$\sigma$ detection limit.

The correction for foreground and background galaxies is performed statistically using the {\it UKIDSS Ultra Deep Survey}\footnote{A detailed description of the survey can be found at:
 \url{http://www.nottingham.ac.uk/astronomy/UDS/}.} (UDS) Data Release 5 field. This field 
consists of a region of $A=0.77\ {\rm {sq. \ deg}}$ in the northern hemisphere with a 5$\sigma$ depth of 21.5 and 22.5 in {\it K} and {\it J} bands, respectively. 
For reasons of homogeneity, the same method is used for carrying out the star-galaxy separation in the background field.

For all galaxies in this field, sky coordinates, total magnitudes and aperture magnitudes 
(used to calculate galaxy colours) are extracted from the UDS database.

\begin{figure*}
\begin{center}
\includegraphics[width=0.48\textwidth,angle=0,clip]{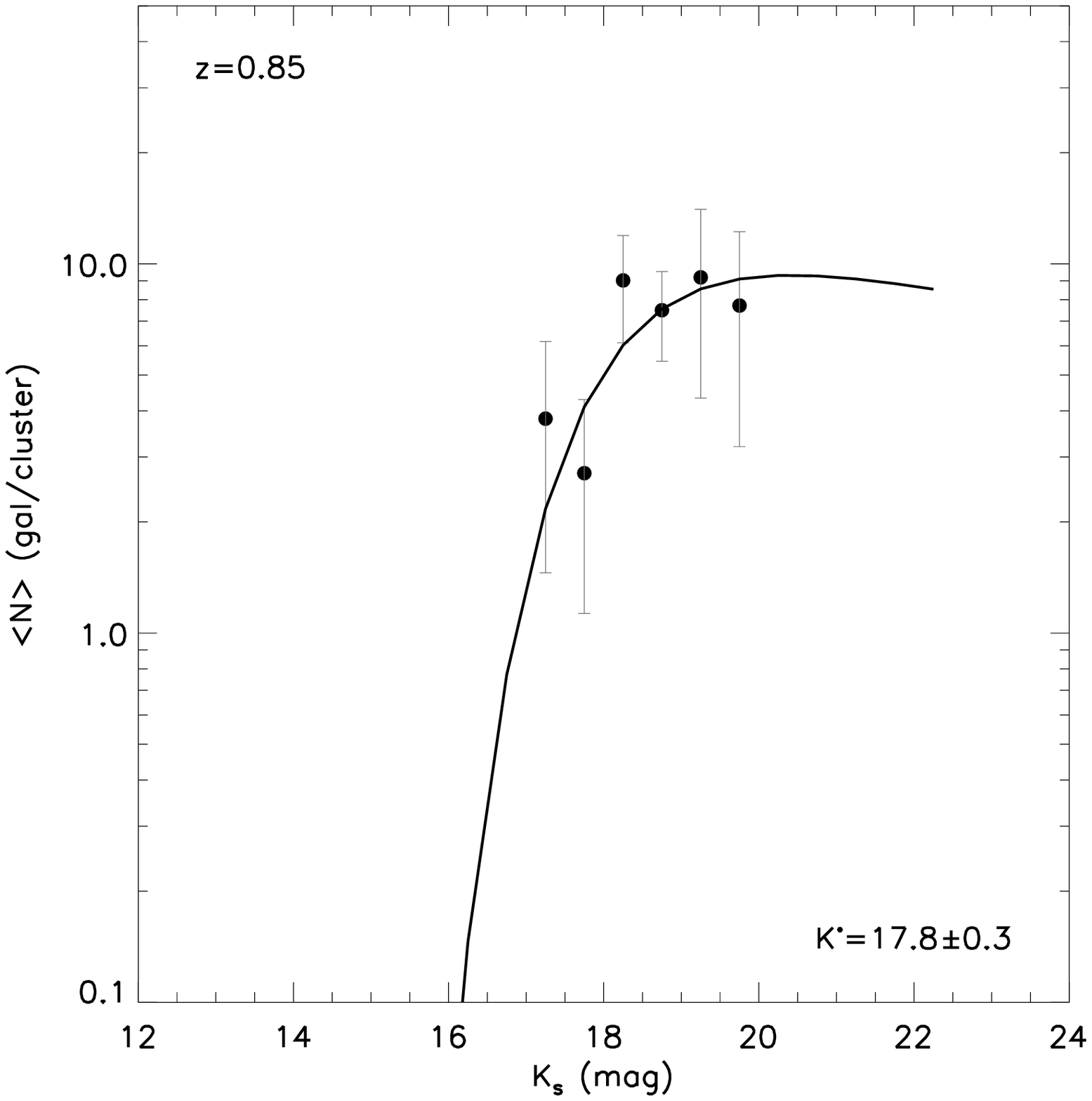} 
\includegraphics[width=0.48\textwidth,angle=0,clip]{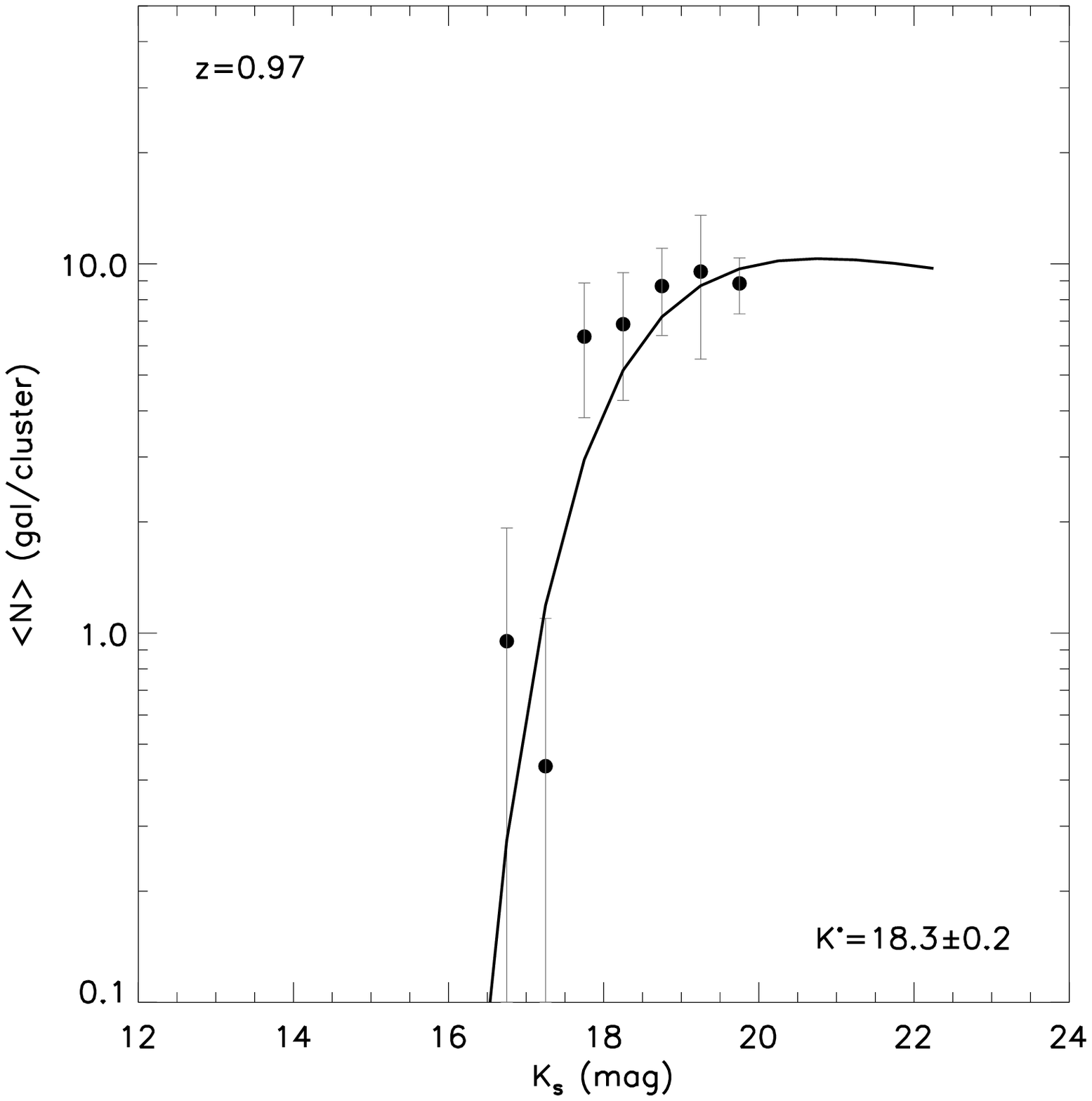} 
\includegraphics[width=0.48\textwidth,angle=0,clip]{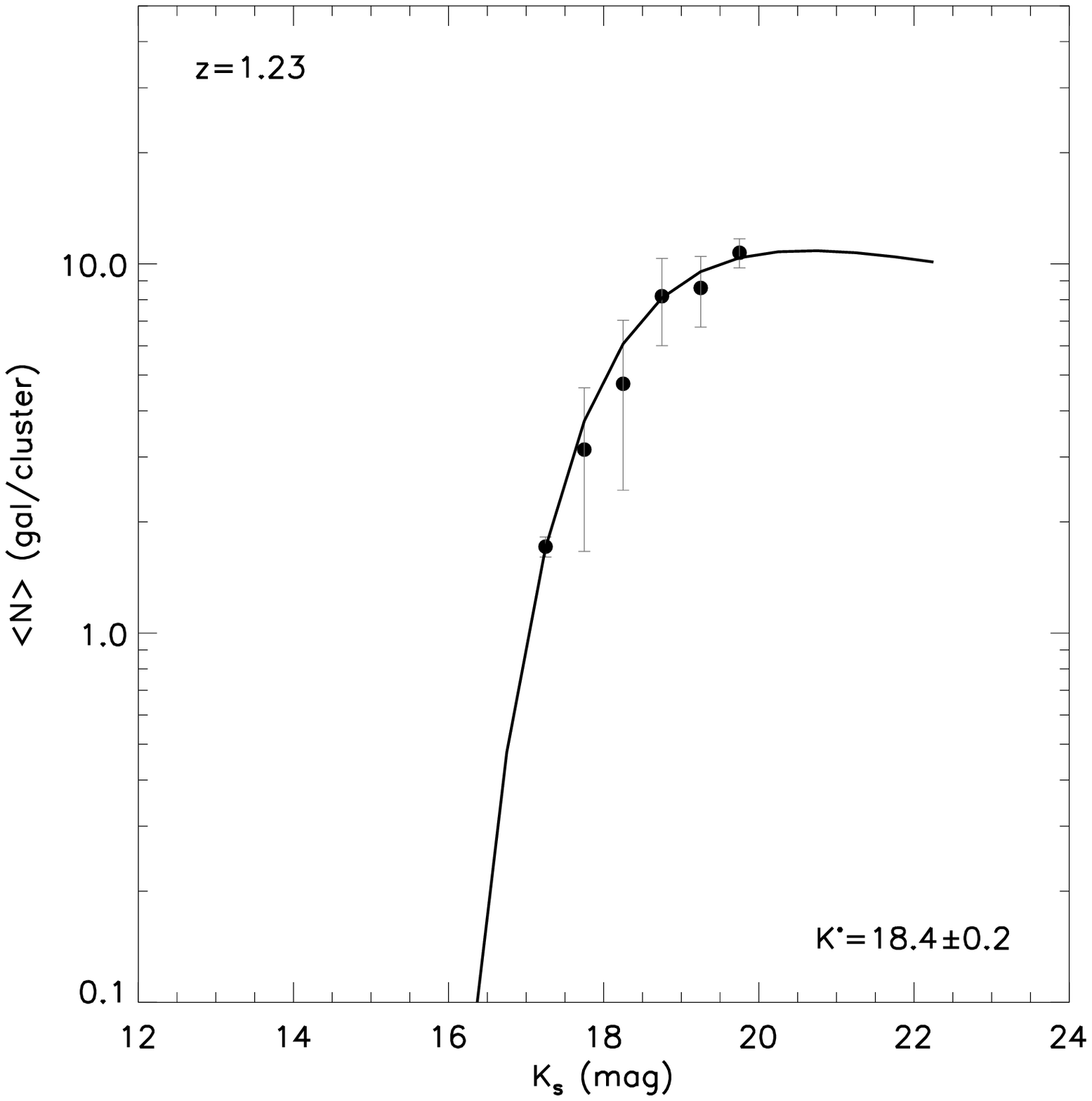} 
\caption{{\it K}-band stacked LFs for our three $z$ bins ($z=$0.85, 0.97, 1.23) with their best-fitting values of ${\it K^{\ast}}$.}.
\label{fig:Figure3}
\end{center}
\end{figure*}

\subsection{{\it K}-Band LF}
\label{subsec:sbsec4.1}
The cluster sample is divided into three redshift bins (median $z$=0.85, 0.97, 1.23) roughly containing the same number of clusters. For each 
cluster, after the stars are removed, we determine the {\it K}-magnitude distribution binned in 0.5-mag intervals. The number of galaxies in each bin $N_{bin}$ is calculated 
using the formula \citep{Ellis-2004}:

\begin{equation}
N_{bin}=N_{cl}-N_{back} \left(\frac{A_{cl}}{A_{back}}\right),
\end{equation}

\noindent where $N_{\rm cl}$ and $A_{\rm cl}$ represent the number of cluster galaxies and the cluster area respectively; while $N_{\rm back}$ and $A_{\rm back}$ are 
the corresponding numbers for the background field.

The corresponding error in each bin interval is found by summing in quadrature the Poissonian error on $N_{\rm cl}$ and the error on the background - which is made up of a  
Poissonian term and a term accounting for galaxy clustering and given by
   
\begin{equation}
\sigma_{N_{back}}=\sqrt{N_{back}} \sqrt{1+\frac{2\pi N A_{\omega} {\theta_{c}}^{2-\delta}}{2-\delta}}.        
\end{equation}

\noindent Here $\theta_{\rm c}$ is the angular radius such that $\Omega=\pi {\theta_{\rm c}}^{2}$, $\Omega$ being the solid angle of the background field, and N is the number 
density of galaxies in the bin considered. The error on $N_{\rm back}$ is then normalised to the projected surface area of the cluster. The parameters $\delta$ and $A_{\rm \omega}$ describe the galaxy angular correlation function such that:

\begin{equation}
\omega(\theta)=A_{\omega} \theta^{\delta}.     
\end{equation}

\noindent We adopt the values $\delta=-0.8$ and $A_{\rm \omega}=(13.49 \pm 1.57)\times 10^{-4}$ taken from \citet{Temporin-2008}, who studied the {\it K}-band angular 
correlation function down to ${\it K}=20.5$, using the {\it VIMOS VLT Deep Survey}\footnote{Detailed information about this survey can be found at: \url{http://www.oamp.fr/virmos/vvds.htm}.}
(VVDS).\\
Similarly to \citet{De-Propris-1999}, we then use the difference in redshift between the cluster redshift and the median redshift of its assigned bin, 
along with the k-correction, to appropriately transform the galaxy magnitudes. In this way, the intervals will align at the median redshift of each bin.\\
The k-correction is calculated by using the model of \citet{Bruzual-2003}.

For each redshift bin, we obtain a stacked luminosity function (Fig. \ref{fig:Figure3}). BCGs are excluded, as they affect the LFs in such a way
that the abundance of very bright galaxies is underestimated when a Schechter function is used for the fitting (e.g., \citealp{Schechter-1976, Christlein-2003}). 
We then fit the stacked LFs obtained in each bin with a Schechter function \citep{Schechter-1976}, fixing the faint end slope to $\alpha=-0.9$, under the 
supposition this parameter does not evolve with cosmic time and solving for both ${\it K}^{\ast}$ and the normalization. This is the value of $\alpha$ measured by \citet{De-Propris-1998} for the {\it K}-band LF of the Coma Cluster 
at $z\sim 0.02$. The choice of fixing $\alpha$ is justified by our photometry only reaching $\sim 1.5\ {\rm mag}$ below $L^{\ast}$. The fitting procedure was 
based on least-squares based on the the Levenburg-Marquardt algorithm. The values of the best fit 
${\it K}^{\ast}$ for the stacked LFs are reported in Table \ref{tab:Table2}, while the best fit LFs (plotted as average galaxy number counts per cluster vs. magnitude) 
are shown in Figure \ref{fig:Figure3}.

Since the value of $\alpha$ measured in the field is $\alpha \sim -1$ \citep{Cirasuolo-2010} we also repeat the fit for the stacked LF 
fixing the slope to $\alpha=-1$. The best fit values of ${\it K^{\ast}}$ obtained (Table \ref{tab:Table2}) are consistent within 1$\sigma$ 
of the ones corresponding to $\alpha=-0.9$. The latter values are plotted in the Hubble diagram in Figure \ref{fig:Figure4}, together with 
the predictions of evolutionary models with different formation $z$ calculated using the \citet{Bruzual-2003} model. The models are 
calculated assuming a simple stellar population (SSP) with a Chabrier initial mass function and solar metallicity. In Figure \ref{fig:Figure4} we also compare our results with the predictions of the SAM by \citet{Bower-2006}, which implements baryon physics, such as AGN feedback from a central black hole and star formation, 
as a SAM bolted onto the MS. Mock clusters were selected from the simulations in seven redshift 
snapshots as described in  Section \ref{sec:section5}. The ${\it K}^{\ast}$ values from the mock clusters are obtained by applying identical selection criteria to the real data, including the same least-squares fitting
procedure.  

\begin{table}
\begin{scriptsize}
\begin{center}
\begin{tabular}{cccccc}
\hline  
$\mathbf {z}$ &  $\mathbf{K^{\ast}}$ &  $\mathbf{K^{\ast}}$  &   $\mathbf{\langle M_{500} \rangle}$   & $\mathbf{\langle N_{500} \rangle}$ & $\mathbf{No. Clusters}$ \\
		&				$ (\alpha=-0.9)$			&		$ (\alpha=-1)$		& ($10^{14}{\rm M}_{\odot}$)   & ({\rm gal})           &\\
		\hline                        			 																					  
0.85    &   17.8$\pm$0.3      			 	    &   17.4$\pm$0.6 				&    	$2.0\pm0.4$	  &		48$\pm$8 							 & 			  5					  \\
0.97    &   18.3$\pm$0.2      			 	    &   18.0$\pm$0.6 				&	$3.6\pm1.2$	  &		56$\pm$7 							 & 			  4					  \\
1.23    &   18.4$\pm$0.2      			 	    &   18.6$\pm$0.3 				&	$3.1\pm0.7$	  &		46$\pm$6 							 & 			  6					  \\
\hline
\end{tabular}
\caption{Here are tabulated, for each redshift bin, the values of median $z$ (col 1), best-fitting ${\it K^{\ast}}$ for $\alpha=-0.9$ (col 2),  
best-fitting ${\it K^{\ast}}$ for $\alpha=-1$ (col 3), average galaxy number within $R_{500}$ (col 4) and number of clusters per $z$ bin (col 5), 
measured for our observed cluster sample.}
\label{tab:Table2}							    
\end{center}
\end{scriptsize}
\end{table}

\subsection{HOD Analysis} 
\label{subsec:sbsec4.2}
Despite the diverse methods for studying the HOD, 
directly counting the number of galaxies within known dark matter haloes remains the most straightforward way to estimate the average galaxy number within a 
halo as a function of halo mass. 
Furthermore, for galaxy clusters their total mass can be estimated via gravitational lensing or, as in this case, using the temperature of the X-ray emitting gas sitting in the cluster's 
gravitational potential well. 
Estimating the halo masses in such a way can constrain quite directly P(N$|$M) at high mass \citep{Berlind-2002}. Furthermore, the 
relation between dark matter and galaxy spatial distributions can be investigated in clusters by means of the radial distribution of their 
galaxies. The latter can be directly probed by their surface density profile. 

As mentioned before in Section \ref{sec:section1}, several studies (e.g. \citealp{Berlind-2002, Kravtsov-2004}) have shown that ${\langle N\rangle}_{\rm M}$ can 
be described as the joint probability that a halo of mass $M$ hosts a central galaxy and that the halo hosts a given number $N_{\rm s}$ of satellite galaxies. Since 
this study investigates the high mass regime of ${\langle N \rangle}_{\rm M}$, where P(N$|\langle N\rangle$) is expected to be Poisson distributed, only satellite cluster galaxies 
are considered; moreover the observed and mock clusters analysed here always have $M_{200}> 10^{14}\ {\rm M_{\odot}}$ where, as described in the Introduction,  ${\langle N \rangle}_{\rm M} \propto M^{\beta}$.  
In addition these clusters (mock and real) are also used to study $c_{\rm g}$ at high redshift. We note here that in order to directly compare our results with those of other workers we transform from $M_{200}$ and $R_{200}$ to the respective values at over density 500. In this paper the mass transformation is always carried out using the standard NFW profile prescription by \citet{Hu-2003} and for the radial scaling we use $M_{\Delta}=(4/3) \pi R_{\Delta}^3 \Delta\rho_c(z)$, where $\Delta$ is the considered overdensity. 

\subsubsection{$N_{500}-M_{500}$ Relation} 
\label{subsubsec:sbsbsec4.2.1} 
After determining the values of ${\it K}^{\ast}$ for each of the three redshift bins, we are able to calculate $\langle N_{500} \rangle$ to a depth of ${\it K}={\it K}^{\ast}+1.5$. 
However, in order to allow us to meaningfully compare our results with other studies at low $z$ , where at least a depth of ${\it K}={\it K}^{\ast}+2$ is reached,
we decided to extrapolate $\langle N_{500} \rangle$ to ${\it K}^{\ast}+2$ (after normalising brighter than ${\it K} = {\it K}^{\ast}+1.5$ and allowing for suitable propagation of  
errors). The values obtained are shown in Table \ref{tab:Table2}. 
We plot these values against $M_{500}$ together with \citet{Lin-2004} data points and best-fit relation at $z \sim 0.06$ in Fig. \ref{fig:Figure5}. 

\begin{figure}
\begin{center}
\includegraphics[width=0.5\textwidth,angle=0,clip]{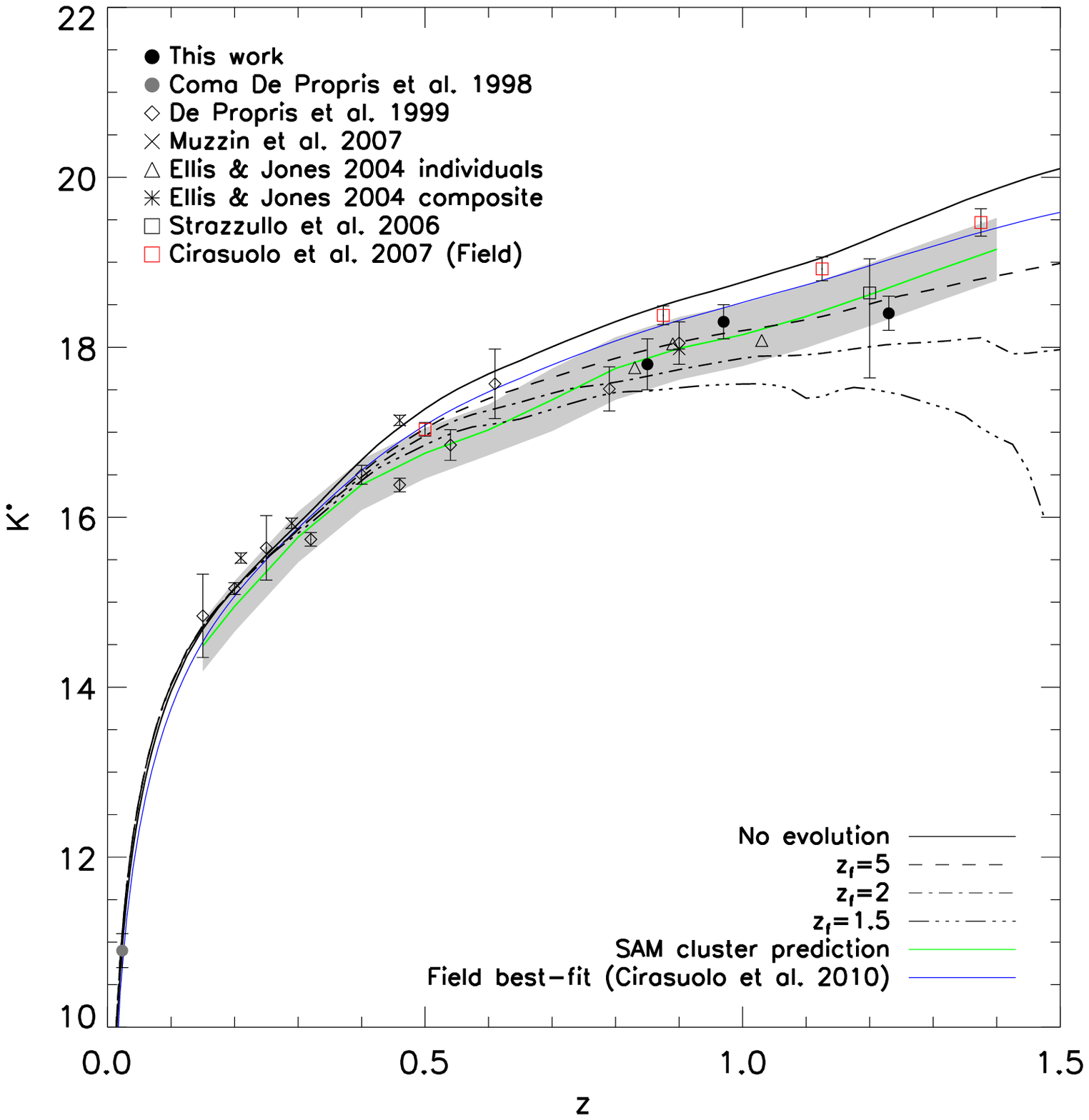} 
\caption{Hubble diagram for ${\it K^{\ast}}$. The data points for the cluster environment are taken from: this work, \citet{De-Propris-1998, De-Propris-1999}, 
\citet{Muzzin-2007a}, \citet{Ellis-2004} and \citet{Strazzullo-2006}. From the field: \citet{Cirasuolo-2007}. Several evolutionary models are also plotted: 
no evolution in  absolute {\it K}-band magnitude (full line), passive evolution with $z_{\rm f}$= 1.5, 2 and 5 (dashed-triple-dotted, dashed-dotted and dashed line respectively) 
and \citet{Bower-2006} SAM predictions (green full line) obtained by using mock clusters (see text). The grey-shaded region shows the typical variation in the evolution of the SAM ${\it K^{\ast}}$ when a constant halo cluster mass is adopted. The scatter is dominated by Poisson statistics.
The parametrization of the observed evolution of the field ${\it K^{\ast}}$
by \citet{Cirasuolo-2010} is also shown (blue full line).}
\label{fig:Figure4}
\end{center}
\end{figure}

\subsubsection{Galaxy Surface Density Profile}
\label{subsubsec:sbsbsec4.2.2}
After selecting only cluster galaxies (BCGs excluded) within $R/R_{500} < 1$ and out to $K=K^{\ast}+1.5$, we stack them all to obtain a 
single radial profile. We then calculate the surface number density normalised per virial area and plot it as a function of virial 
fraction ($R/R_{200}$). In this way we are able to obtain a stacked radial number density profile and to study the mean concentration of 
cluster galaxies at $z\sim 1$. In order to estimate $c_{\rm g}$ we fit a NFW density profile to the data. 
The three-dimensional NFW profile has the following form:

\begin{equation}
\frac{\rho(r)}{\rho_{c}}=\frac{\delta_{c}}{(r/r_{s}){(1+r/r_{s})}^{2}},
\end{equation}

\noindent where $\rho_{\rm c}=3{H_{0}}^{2}/8 \pi G$ is the critical density for closure, $\delta_{\rm c}$ is a characteristic density contrast and 
$r_{\rm s}$ is a scale radius. Since we consider a density contrast of $\Delta=200$, $\delta_{\rm c}$ and $r_{\rm s}$ take the following form,

\begin{equation}
\delta_{c}=\frac{(200/3)c^{3}}{[\ln(1+c)-c/(1+c)]},
\end{equation}

\begin{equation}
r_{s}=r_{200}/c,
\end{equation}

\noindent where $c$ is the concentration parameter of the profile. The surface density is then obtained by projecting the three-dimensional 
profile along the line of sight from negative to positive infinity and, following \citet{Bartelmann-1996}, can be written as

\begin{equation}
\label{eq:NFW_proj}
\Sigma(x)=\frac{2\rho_{s}r_{s}}{x^{2}-1}f(x),
\end{equation}
 
\noindent where $\rho_{\rm s}=\delta_{\rm c}\rho_{\rm c}$, $x=c r/r_{200}$ and $f(x)$ is given by:

\begin{equation}
f(x)=\Bigg\{
\begin{array}{ll}
1-\frac{2}{\sqrt{x^{2}-1}}\arctan\frac{\sqrt{x-1}}{\sqrt{x+1}} & (x>1),\\
1-\frac{2}{\sqrt{1-x^{2}}}{\rm{arctanh}} \frac{\sqrt{1-x}}{\sqrt{1+x}} & (x<1),\\
0 &  (x=1).
\end{array} 
\end{equation}

\begin{figure}
\begin{center}
\includegraphics[width=0.5\textwidth,angle=0,clip]{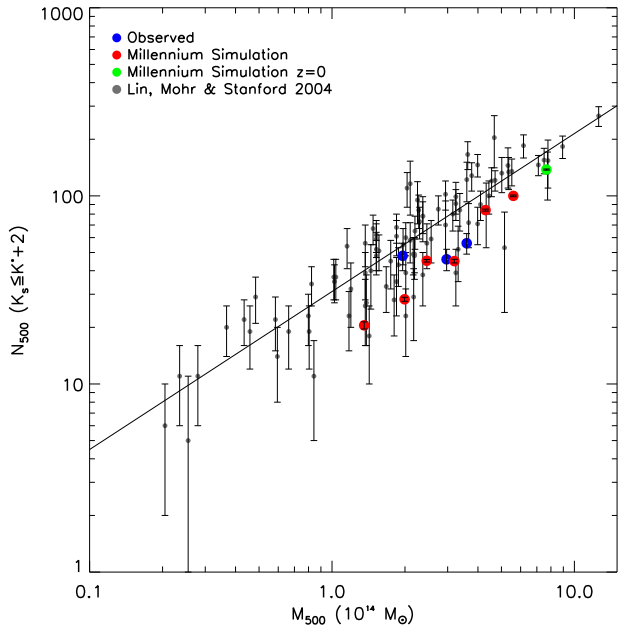} 
\caption{$N_{500}$ against $M_{500}$. Our mean values are reported as blue full dots (observed sample), red full dots (mock clusters with 
$z>0$) and green full dot (mock clusters at $z=0$). Note that the observed data point at $z=1.23$ has been slightly shifted towards lower masses to make it visible. The data points of \citet{Lin-2004} --plotted as grey full dots--  together with their best-fitting relation are 
shown. We note that,  since \citet{Lin-2004} values are measured over clusters at  $z\sim 0.06$, only the average value for $N_{500}$ at $z=0$ (green full dot) can be directly compared with them. Note our simulation-based measures are average values measured over the 100 most massive clusters in each snapshot (see text).} 
\label{fig:Figure5}
\end{center}
\end{figure}

\noindent The limited FOV of {\it MOIRCS} does not allow us to observe galaxies far enough from the cluster to be able to determine the 
background level by investigating regions of the sky dominated by field galaxies. For this reason, and because we do not have spectroscopic 
redshifts for our cluster galaxies, we can not directly background subtract our stacked number density profile. Hence we start with the 
methodology of \citet{Lin-2004} and carry out a maximum likelihood fit to the 
projected radial density profile with the supposition of having a constant background. The observed profile is binned very finely in virial 
fraction ($R/R_{200}$) so that in every bin, the density within the differential area $dS$ could be considered constant. By doing this we 
can estimate the probability of obtaining the measured galaxy counts within $dS$ for a specific prediction
($N_{\rm model}$) given by the model. This situation is well represented by a Poisson probability distribution with mean 
$\lambda=N_{\rm model}$. Our model is simply given by the sum of the cluster contribution to the projected density and the contribution of the 
background:

\begin{equation}
\Sigma=\Sigma_{cl}+\Sigma_{back},
\label{eq:eq9}
\end{equation}      

\noindent where $\Sigma_{\rm cl}$ is given by eq. \ref{eq:NFW_proj} and $\Sigma_{\rm back}$ is a constant. Specifically, the quantity

\begin{equation}
-\log \mathcal{L}=-\sum_{\rm i}^{N_{\rm bin}} \log \big[\frac{{\rm e}^{-\lambda_{\rm i}} {\lambda_{\rm i}}^{N_{\rm obs_{\rm i}}}}{N_{\rm obs_{\rm i}}!} \big],
\label{eq:eq10}
\end{equation}

\begin{figure}
\begin{center}
\includegraphics[width=0.5\textwidth,angle=0,clip]{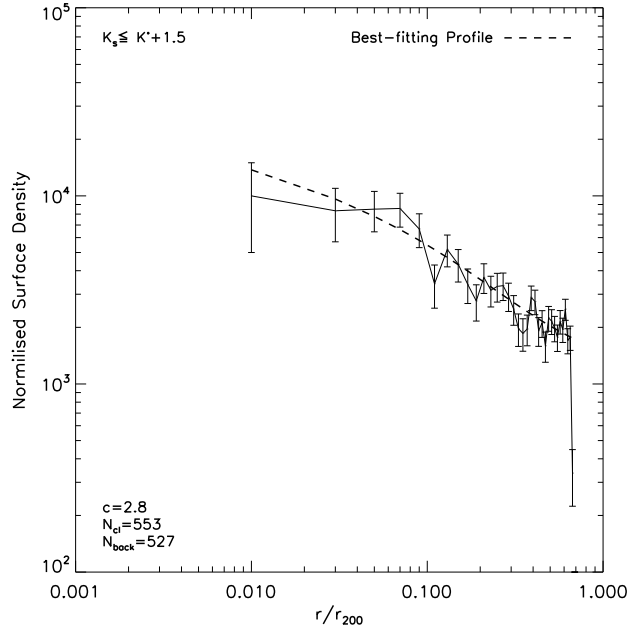} 
\caption{Stacked radial surface number density profile for our observed cluster sample. Galaxies within $R_{500}$ and out to ${\it K}=
{\it K^{\ast}}+1.5$ only are used. The best-fitting profile is shown as a dashed line.}
\label{fig:Figure6}
\end{center}
\end{figure}

\noindent is minimised, where $N_{\rm bin}$ is the number of bins used, $N_{\rm obs_{\rm i}}$ is the observed number counts in the differential area $dS_{\rm i}$ and $\lambda_{\rm i}=\Sigma(x_{\rm i})dS_{\rm i}$ is
the number of galaxies predicted by the model (cluster galaxies plus background, $N_{\rm model}$) in the same differential area as a function of $\rho_{\rm s}$, $c=c_{\rm g}$ and 
$\Sigma_{\rm back}$. In the fitting routine we allow the parameters $\rho_{\rm s}$, $c_{\rm g}$ and $\Sigma_{\rm back}$ to vary. The advantage of this method is that the likelihood 
involves a proper sum of Poisson probabilities (in this respect our method improves on the one used by \citealp{Lin-2004}) and that, unlike methods based on the $\chi^{2}$, the results are relatively independent of the binning. Figure \ref{fig:Figure6} shows the stacked radial density 
profile out to $\sim 0.65 R_{200}$ corresponding to the $R_{500}$ radius considered for each cluster. The determined best fitting galaxy concentration parameter is 
$c_{\rm g}=2.8_{-0.8}^{+1}$, with the uncertainties derived from the relation $\Delta \chi^{2}=-2\Delta\log \mathcal{L}$. The value obtained is consistent with those found by \citet{Carlberg-1997}, 
\citet{van-der-Marel-2000}, \citet{Lin-2004} and \citet{Biviano-2010} for clusters at low and intermediate redshifts ($c_{\rm g}$= 3.7, 4.2, 2.9 and 4.0 at $z\sim 0.3$, 0.3, 0.1 and 0.55, respectively).
We must point out though, that the magnitude limit used in these studies does not always coincide with ours.

The background density obtained amounts to around $\Sigma_{\rm back}\sim 8$ per cent of the peak density of the stacked profile or $\simeq 49$ per cent of total galaxies (527 out of 1080). We tested the reliability of these estimates by analysing the radial number density profile out beyond $R_{200}$. This is done by  using those clusters in the sample (3 out of 15) whose 
$R_{200}$ is small enough to allow us to estimate the background out to at least $1.4 R_{200}$. For these clusters  the background level is evaluated by averaging the values in the outermost bins ($R/R_{200}>1$) 
of the radial profile. The values obtained vary between 4 and 12 per cent of the density peak value, corresponding to $30-50$ per cent of the total galaxy number out to the same radial distance as used in the stacked profile ($R\simeq 0.65 R_{200}$). This result is in relatively good agreement with the background level estimated with the likelihood. 

\begin{table*}
\begin{center}
\begin{tabular}{lcccccc}
\hline  
 {\bf Redshift} & $\mathbf{ \beta}$      & $\mathbf{\langle M_{500} \rangle}$  	&   $\mathbf{\langle N_{500} \rangle}$  & 	$\mathbf{c_{\rm g}}$        & $\mathbf{c_{\rm dm}}$  &    $\mathbf{\Delta \log c}$\\  
                      &  &   ($10^{14}$M$_{\odot}$)   &  ({\rm gal})       &      &        &\\										  
\hline         																					   										                                                                                                                                       
  0 	   &  $0.82 \pm0.01 \ (3769)$ 	 & 			    		$7.6\pm0.3$				&  	 138$\pm$1  		   				 & 	$6.30_{-0.36}^{+0.39}$  & 3.74					&    $0.23_{-0.02}^{+0.03}$  \\       								 
  0.21 	&  	  $0.76 \pm0.01 \ (2752)$		 & 			 $5.6\pm0.2$ 						&  	 100$\pm$1  		   				 & 	$5.31_{-0.35}^{+0.39}$  &   - 					&  						 - 	  \\ 			      									 
  0.41    &       $0.82 \pm0.02 \ (1953)$ 		 & 			 $4.3\pm0.1$   						&  	  84$\pm$1  		   				 & 	$5.05_{-0.38}^{+0.41}$  &   - 					&  						 - 	  \\  			      									 
  0.62	   &	   	 								   $0.86\pm0.02 \ (1289)$ 		 & 			 $3.2\pm0.1$						&  	  45$\pm$1  		   				 & 	$4.90_{-0.47}^{+0.52}$  & 3.48					&    $0.15_{-0.04}^{+0.05}$  \\        							 
  0.83	   &	    	 								   $0.74\pm0.03 \ (831)$ 		 & 			 $2.4\pm0.1$   						&  	45$\pm$1 	      				 & 	$3.85_{-0.40}^{+0.45}$  &   - 					&  					  -		  \\   			      										

  0.99	   &	    	 								   $0.78\pm0.04 \ (541)$ 		 & 			 $2.0\pm0.1$   						&  	28$\pm$1 	      				 & 	$3.70_{-0.44}^{+0.50}$  & 3.66					&    $0.00_{-0.05}^{+0.06}$  \\       								 
  1.27	   &	    	 								   $0.81\pm0.08 \ (233)$		 & 			 $1.40\pm0.05$ 						&  	21$\pm$1 	      				 & 	$3.77_{-0.51}^{+0.58}$  &   - 					&  					  -		  \\  			      										      							 
\hline
\end{tabular}
\caption{Properties of the mock cluster sample. Col1: redshift of the snapshot bin; col2: the slope $\beta$ of the $N_{500}-M_{500}$ relation (HOD power-law index) for all the mock clusters in the 
snapshot with the number of clusters shown in parenthesis - note that all other estimates from the mock sample in this table use the 100 most massive cluster haloes at each redshift snapshot; col3:
the average mass within the radius corresponding to $\Delta=500$; col4: the average galaxy number within $R_{500}$ brighter than $K^{\ast}+2$ (these values do not include the systematic
correction of 10 per cent discussed in Section \ref{subsec:sbsec5.2}); col5: the average galaxy concentration parameter for the mock clusters; col6:  the average dark matter concentration parameter 
as measured from the $c_{\rm dm}(M,z)$ relation of \citet{Gao-2008}, not available for all z; col9: $\Delta \log c= \log(c_{\rm g}/c_{\rm dm})$, see text.}
\label{tab:Table3}							    
\end{center}
\end{table*}

\section{Analysis of Mock Cluster Samples}
\label{sec:section5}
\subsection{Cluster Halo Selection}
\label{subsec:sbsec5.1}
The MS\footnote{Detailed information about the simulation can be found at: \url{http://www.mpa-garching.mpg.de/millennium/}. A detailed description of the data base can be
found in \citet{Lemson-2006}.} is a large N-body simulation of the standard $\Lambda CDM$ cosmogony. It follows $2160^{3}$ particles, 
each of mass $8.6\times 10^{8}\ {\rm h^{-1}\ M_{\odot}}$, in a cubic box of $500\ {\rm h^{-1}\ Mpc}$ on a side from redshift $z=127$  to the present. The cosmological parameters 
were chosen to be consistent with a combined analysis of the 2dFGRS \citep{Colless-2001, Percival-2001} and first-year {\it Wilkinson Microwave Anisotropy Probe (WMAP)} data 
\citep{Spergel-2003} within the concordance cosmology.

By scanning the entire volume of the simulations we extract all haloes above a minimum mass of $M_{200}> 10^{14}\ {\rm M_{\odot}}$, selected in seven snapshots
corresponding to redshifts $z=0$, 0.21, 0.41, 0.62, 0.83, 0.99 and 1.27, respectively. 

The purpose of the simulations is to follow the evolution of the observed clusters, which do not constitute a statistical sample but are highly biased to the most X-ray luminous systems and therefore the most massive clusters at $z>0.8$. The finite volume of the MS means that the number of massive clusters is rare, so for the study of the evolution of ${\it K}^{\ast}$, $\langle N_{500} \rangle$ and $c_{\rm g}$, we restrict our analyses to the most massive cluster haloes in each snapshot of the simulation, since they are thought of as equivalent to the brightest X-ray systems. This approach follows that of \citet{De-Lucia-Blaizot-2007} who used the 125 most massive clusters in similar snapshot intervals from the MS to investigate the evolution of BCGs in X-ray selected luminous clusters (see Sect. 6 of their paper). Selecting the most massive 100 clusters per snapshot we can reproduce both the observed redshift and mass distributions of our high-redshift clusters reasonably well; for the mocks:     
 $0.83<z<1.27$, $1<M_{500}<7.7\times 10^{14}\ {\rm M_{\odot}}$ and median $M_{500}=1.8 \times 10^{14}\ {\rm M_{\odot}}$, and for the observations: $0.81<z<1.46$, $0.4<M_{500}<6.2\times 10^{14}\ {\rm M_{\odot}}$ (next least massive cluster is at $0.9\times 10^{14}\ {\rm M_{\odot}}$, so only one observed cluster lies significantly outside the mock mass range), and median $M_{500}=2 \times 10^{14}\ {\rm M_{\odot}}$. Details of the mock clusters used are given in Table \ref{tab:Table3}.
 
One consequence of selecting the most massive haloes per snapshot is that the mass range evolves with cosmic time (as in the analysis of  \citealp{De-Lucia-Blaizot-2007}), so to investigate whether our results are biased by this mass evolution we also present results using other mock cluster samples selected with a constant average mass and a constant mass range  over cosmic time. Results of these analyses are reported in Section \ref{subsubsec:subsubsec6.2.3}. The galaxies populating the haloes used are extracted from the SAM (see also Section \ref{subsec:sbsec4.1}) galaxy catalogue by \citet{Bower-2006}. \\

\subsection{Satellite Galaxies' Power-law Index \& Occupation Number} 
\label{subsec:sbsec5.2}
Once the galaxy spatial coordinates and {\it K}-band magnitude are extracted for the totality of haloes contained in each redshift bin, we select galaxies with projected radial distances within $R_{500}$ 
for each mock catalogue and carry 
out individual fits to obtain individual LFs in the same way as described for the observations in Section \ref{subsec:sbsec4.1}. We point out that in order to reproduce observations the cut in $R_{500}$ is not applied 
along the line of sight and that galaxies are assigned to each halo by using the unique halo IDs, which allow us to univocally associate galaxies to the haloes they belong to. However, on larger scales the lack of a background in the mocks means that the galaxy selection differs from the observed clusters, for which galaxies are counted within a cylinder of projected radius $R_{500}$ and a background correction applied using galaxy counts (see Section \ref{subsec:sbsec4.1}). In principle line-of sight contamination from galaxy clustering on scales extending beyond the virial radius of the clusters could affect our richness and concentration values by introducing an offset between measured and mock estimates. Therefore to check this we inspect the number of non-cluster galaxies at $z=1$ brighter than $K^{\ast}+2$, with physical and projected distances from mock halo central galaxies of $R_{500}<R<3\ {\rm Mpc}$ and $R<{R_{500}}$ respectively. We find that $N_{500}$ values for the mock clusters are $\simeq10$ per cent lower on average compared to those observed due to this contamination. This is less than the individual Poisson error on $N_{500}$ for each cluster. From an observational perspective these results are broadly consistent with \citet{De-Filippis-2011} who compared what effect local and global background corrections have on the counts in clusters at low redshift.

The radial distances for the mock sample are calculated with respect to the central galaxy of the halo, which is then excluded from the fitting process. Once the best-fitting 
values of ${\it K}^{\ast}$ are derived, we are able to evaluate $N_{500}$ (out to ${\it K}={\it K}^{\ast}+2$) for each mock cluster and fit a power law of the form:

\begin{equation}
\log N_{500}=\alpha + \beta \log M_{500}. 
\label{eq:eq11}
\end{equation} 

\noindent As described in Section \ref{subsec:sbsec4.2}, the choice for using such a simple relation for fitting the data points in our plot, is justified by the fact we analyse only satellite galaxies within haloes of mass  ($M_{200}> 10^{14}\ {\rm M_{\odot}}$)
significantly higher than the cutoff halo mass ($\gtrsim 10^{12}\ {\rm M_{\odot}}$) introduced in the models (e.g., \citealp{Kravtsov-2004, Zheng-Coil-2007}). 

In this way we are able to investigate the slope of the $N_{500}-M_{500}$ relation and compare the obtained value from the simulations, with those obtained in the literature. 
This process is carried out for each of the aforementioned simulation snapshots, allowing us to study the evolution of the slope $\beta$ of the $N_{500}-M_{500}$ relation with 
cosmic time.The best-fit values of $\beta$ are reported in Table \ref{tab:Table3}.\\

In order to investigate the evolution of the mean number of satellite galaxies $\langle N_{500}\rangle$ per halo mass with cosmic time, we restrict ourselves to the average values 
($\langle N_{500}\rangle$;  $\langle M_{500}\rangle$) for the 100 most massive clusters in each snapshot and carry out the same process utilized for the observed sample, now based on the stacked LF of the mock clusters. The values of $\langle N_{500}\rangle$ obtained are reported in Table \ref{tab:Table3} and plotted against $M_{500}$ in Fig. \ref{fig:Figure5}. 
 
\begin{figure}
\begin{center}
\includegraphics[width=0.5\textwidth,angle=0,clip]{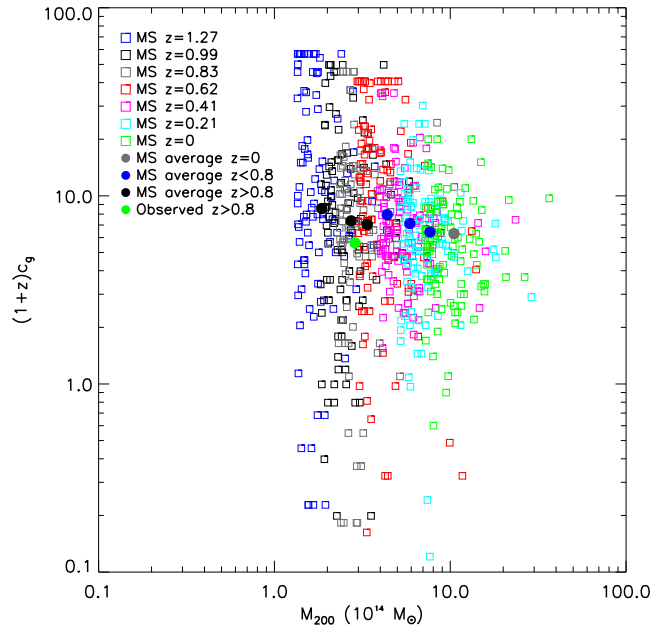} 
\caption{$(1+z)c_{\rm g}$ against $M_{200}$. Measures of $c_{\rm g}$ are derived individually for the 100 most massive mock clusters in each 
of the considered snapshots. They are plotted as: blue squares ($z=1.27$), black squares ($z=0.99$), grey squares ($z=0.83$), red 
squares ($z=0.62$), magenta squares ($z=0.41$), cyan squares ($z=0.21$), green squares ($z=0$). Average values are plotted as well: black 
full dots (mock clusters with $z>0.8$), blue full dots (mock clusters with $z<0.8$), grey full dot (mock clusters with $z=0$), green full 
dot (observed clusters).}
\label{fig:Figure7}
\end{center}
\end{figure}

\subsection{Concentration Parameter} 
\label{subsec:sbsec5.3}
For each set of mock clusters, the value of $c_{\rm g}$  is also calculated using the same maximum-likelihood fitting procedure used for the observed clusters. 
The model used for the fitting procedure is similar to that described in Section \ref{subsubsec:sbsbsec4.2.2}, but with $\Sigma_{back}=0$ (see Eq. \ref{eq:eq9}), since
the mock clusters do not suffer from contamination by galaxies not belonging to them due to the unambiguous way they are assigned to the mock haloes.
The fitting procedure is applied to all the clusters individually (see Fig. \ref{fig:Figure7}) and also applied to the stacked projected number 
density profiles,  which are first normalized to the virial area within $R_{200}$ (see Table \ref{tab:Table3}).  Although at a given halo 
mass, the concentration parameters are commonly assumed to be lognormally distributed (e.g., \citealp{Jing-2000, Neto-2007, 
Comerford-2007}), we prefer in this work to use the average values of $c_{\rm g}$ from the stacked profiles to compare with observations
because the stacking procedure has the advantage of erasing individual deviations, usually related to the presence 
of substructure \citep{Gao-2008}. These deviations may contribute to the already considerable intrinsic scatter (\citealp{Neto-2007} treated 
this issue in details) in the parameters fitted to individual profiles, possibly masking underlying trends in the data. It is worth pointing 
out that for the stacked profiles only the radial range $0.05<R/R_{200}\lesssim 0.65$ is used in the fitting process in order to avoid the 
innermost radial region where the NFW model fails to correctly reproduce the density profile, as shown in several 
studies in the literature (e.g., \citealp{Navarro-2004, Prada-2006} and \citealp{Gao-2008}). The best fit values of the average $c_{\rm g}$ are 
reported in Table \ref{tab:Table3}.

\section{Results}
\label{sec:section6}

\subsection{Evolution of $K^{\ast}$}
\label{subsec:subsec6.1}
The ${\it K}^{\ast}$ values for our 15 clusters in the 3 redshift bins $0.85, 0.97, 1.23$ are shown in Table \ref{tab:Table2} and compared to other results in Figure \ref{fig:Figure4}. Our values of $K^{\ast}$ 
are consistent with other observations at similar wavelengths of other high-redshift clusters (e.g., \citealp{De-Propris-1999, Ellis-2004, Strazzullo-2006, Andreon-2006, Muzzin-2007a}). From a direct comparison with the $K^{\ast}$ values predicted 
from our analysis of the mock clusters in the SAMs of \citet{Bower-2006}, we find good agreement: indicating, for the first time, that the SAMs do a reasonable job of predicting $K^{\ast}$ evolution in rich clusters
at redshifts $z\geq1$. However, our data cannot distinguish between the mass-assembly SAMs and pure passively evolving SSP models with formation redshift $z_f\simeq5$, implying that red, old massive galaxies are already ubiquitous in rich clusters at $z\simeq1.5$.

\begin{figure}
\begin{center}
\includegraphics[width=0.5\textwidth,angle=0,clip]{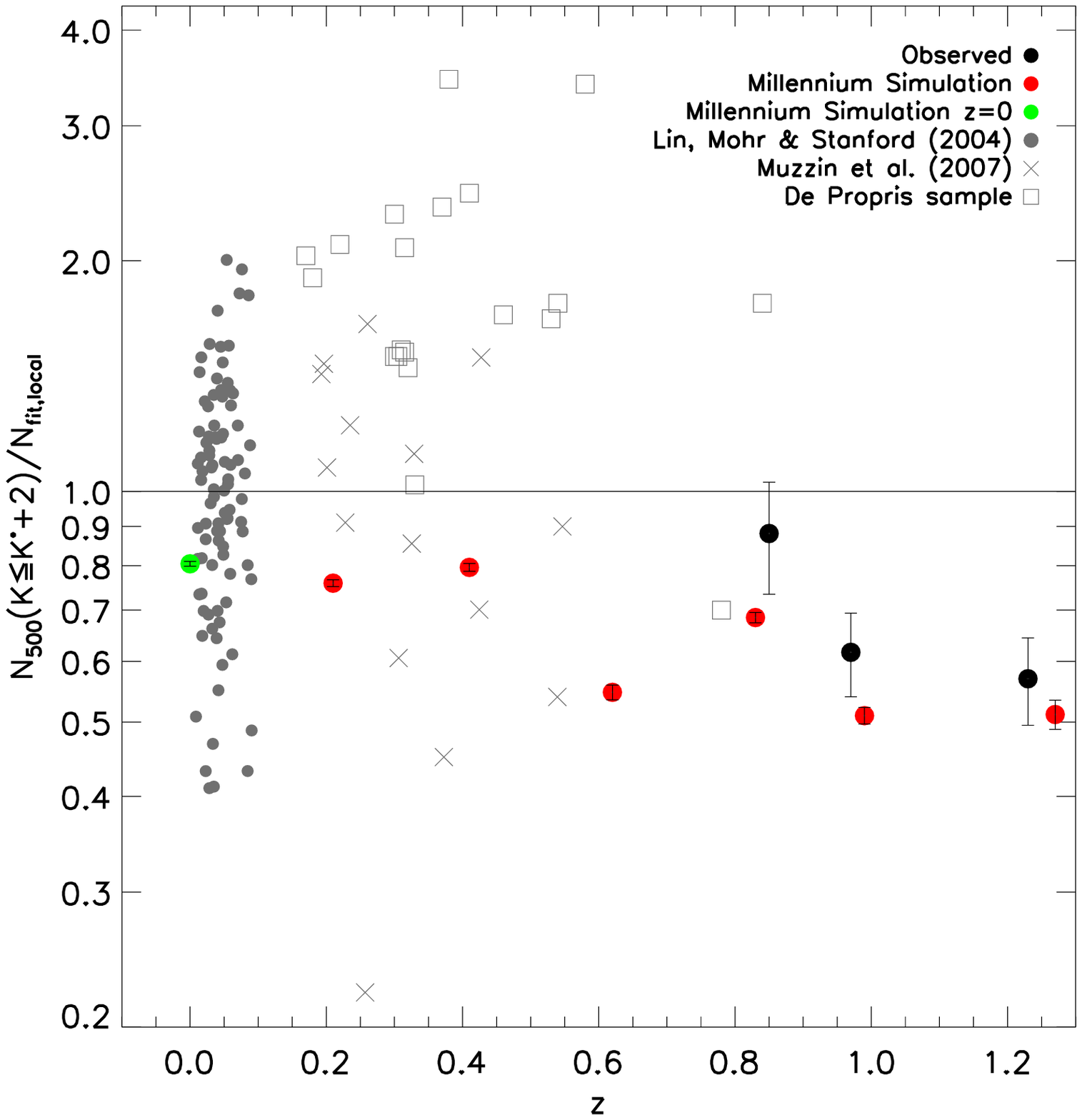} 
\caption{Ratio of measured $N_{500}$ over the prediciton of the low-$z$ best-fitting relation of \citet{Lin-2004}, against $z$: observed clusters 
(black full dots), mock clusters with $z>0$ (red full dots), mock clusters with $z=0$ (green full dot), \citet{Lin-2004} data points 
(grey full dots), \citet{De-Propris-1999} sample as reanlaysed by \citet{Lin-2004} (empty squares), \citet{Muzzin-2007b} (crosses, note these values are extrapolated out to ${\it K}^{\ast}+3$). Note our simulations based measures 
are average values measured over the 100 most massive clusters in each snapshot.}
\label{fig:Figure8}
\end{center}
\end{figure}

\begin{figure}
\begin{center}
\includegraphics[width=0.5\textwidth,angle=0,clip]{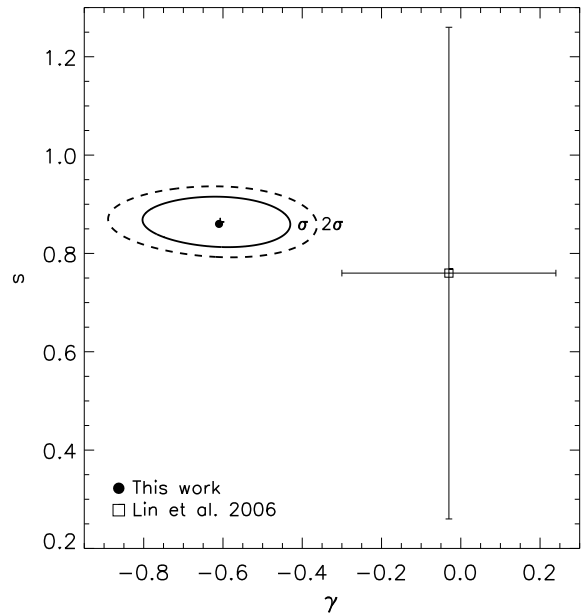} 
\caption{Joint 68 and 95 per cent confidence regions for $\gamma$ and $s$ (see eqn. \ref{eq:eq12}). Black full point shows the best-fitting value for $\gamma$ and $s$ (-0.61, 0.86), while the cross stands for the center of the contours obtained during the marginalisation. Empty square represents the best-fitting values of $\gamma$ and $s$ (plotted together with 
their 1 $\sigma$ errors) by \citet{Lin-2006} when including also the low-$z$ cluster sample used here, in addition to their intermediate-$z$ cluster sample.}
\label{fig:Figure9}
\end{center}
\end{figure}

\begin{figure}
\begin{center}
\includegraphics[width=0.5\textwidth,angle=0,clip]{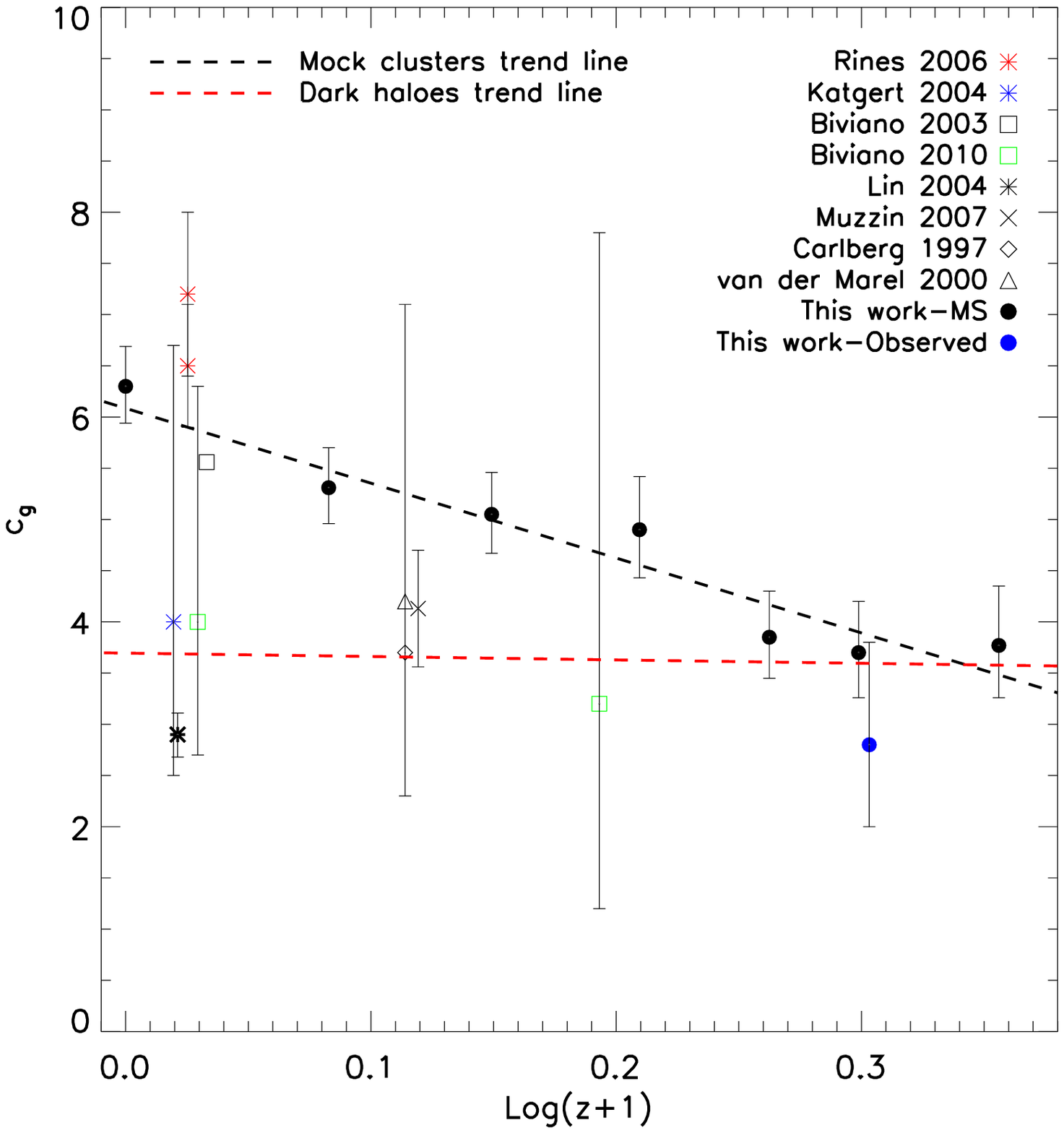} 
\caption{Average galaxy concentration parameter against $\log (z+1)$: this work (black full dots for mock clusters and blue full dot for 
our observed sample), \citet{Rines-2006} (red asterisks), \citet{Katgert-2004} (blue asterisk, slightly shifted towards lower $z$ to improve data point visibility), \citet{Biviano-2003} (black empty square), 
\citet{Biviano-2010} (green empty squares), \citet{Lin-2004} (black asterisk), \citet{Muzzin-2007a} (black crosses, slightly shifted towards higher $z$ for clarity), \citet{Carlberg-1997} (black empty rhombus), 
\citet{van-der-Marel-2000} (black empty triangle). Note for mock clusters, the 100 most massive clusters in each snapshot are used for 
this measure. Trend lines for the simulation-based average values of $c_{\rm g}$ measured in this work (black dashed line) and for the predictions 
of \citet{Gao-2008} derived for dark matter haloes with the same halo masses as our mock clusters (red dashed line) are also plotted.}
\label{fig:Figure10}
\end{center}
\end{figure}

\subsection{HOD}
\label{subsec:subsec6.2}

\subsubsection{Observations}
\label{subsubsec:subsubsec6.2.1}
The observed $ \langle N_{500} \rangle $ values measured from our clusters ($0.8\lesssim z \lesssim 1.5$) are shown in Fig. \ref{fig:Figure5}, where the 
local data from 93 clusters from \citet{Lin-2004} are also plotted along with their best-fit $N_{500}-M_{500}$ line. At all redshifts our points lie below the local fit. In Figure \ref{fig:Figure8} we show the variation of $ \langle N_{500} \rangle $ with redshift. Here the values of 
$N_{500}$ have been normalised to the low-$z$ best-fit relation shown in Figure \ref{fig:Figure5} using the relation showed in Table 1 (second row) in \citealp{Lin-2004} (see also Fig. 9 in their paper).
The values measured for our clusters have  $0.35\lesssim N_{500}/N_{\rm fit, local} \lesssim 0.8$. This result is in stark contrast to the results of the re-analysis by Lin and collaborators of the \citet{De-Propris-1999} intermediate-$z$ cluster sample, also shown in Figure \ref{fig:Figure8}, which indicate normalised $N_{500}$ values typically between $1-4$ for clusters in the range  $0.2 < z < 0.8$. However, our results 
are consistent with those of \citet{Muzzin-2007b}, who carried out a similar study to ours on 15 CNOC1 clusters at $0.19<z<0.55$, and of \citet{Andreon-2008}, who found evidence
of a possible break of the cluster scaling relations at $z\sim 1$.

To quantify the dependence of $N_{500}$ on $z$ and cluster mass in our data, we fit the $N_{500}$ data to the relation used by \citet{Lin-2006}, and given by

\begin{equation}
N(M,z)=N_{0}(1+z)^{\gamma}(M/M_{0})^{s}, 
\label{eq:eq12}
\end{equation}

\noindent where $N_{0}$ and $M_{0}$ (set at $10^{14.3}\ {\rm {M_{\odot}}}$ as in \citealp{Lin-2006}) are the normalization factors of the relation. In the fit we use our binned results for the 15 high redshift clusters and the 93 low redshift data points of \citealp{Lin-2004}.
The best-fitting values are 
$N_{0}=53\pm1$, $\gamma=-0.61^{+0.18}_{-0.20}$ and $s=0.86\pm0.05$. Figure \ref{fig:Figure9} shows the contours corresponding to the 68 and 95 per cent confidence regions for $\gamma$ and $s$. The quoted errors on $\gamma$ and $s$ are calculated by projecting the 68 per cent contour onto  $\gamma$ and $s$ axes respectively.

Comparable measurements at our redshifts are those of \citealp{Lin-2006} who analysed 27 clusters with $0<z<0.9$. Our $s$ value agrees reasonably well with their estimate ($s\simeq0.8 \pm 0.5$) although evidence for an evolutionary trend in their data  is significantly weaker ($\gamma=-0.03\pm0.27$). In fact, our value of $\gamma$ indicates a significant trend with $z$ at $\sim3 \sigma$ level; although with only 3 points at $z\geq0.85$ more data and better redshift sampling are required to establish accurate trends. We are in fact also consistent with \citet{Lin-2006} at $\sim 2 \sigma$ level. Unfortunately we cannot test our results by
using  \citet{Lin-2006} data points to repeat our fit, since their values are calculated within $R_{2000}$.\\

Turning to the study of the concentration parameter, the value of $c_{\rm g}=2.8_{-0.8}^{+1}$ found for the observed sample at $z\sim 1$, is consistent 
within $1\sigma$ with the one of \citet{Lin-2004} at $z\sim 0.06$ ($c_{\rm g}=2.90_{-0.22}^{+0.21}$). However, the latter value is quite different from the ones measured at similar redshifts (e.g., \citealp{Rines-2006}).
In Figure \ref{fig:Figure10} the current measure of $c_{\rm g}$ is plotted, together with the values found in the literature, as a function of $\log (z+1)$. 

A comparison with the predictions of \citet{Gao-2008} for dark matter haloes (see Section \ref{subsubsec:subsubsec6.2.2}) at the same redshift as the median of the clusters studied here 
is also carried out, showing that our value of $c_{\rm g}=2.8_{-0.8}^{+1}$ is consistent with this prediction ($c_{\rm dm}=3.55$) (see Fig.  \ref{fig:Figure10}).   

\subsubsection{Observations \& Simulations: Most Massive Clusters}
\label{subsubsec:subsubsec6.2.2}
Turning to the results obtained for the simulations, the values of $N_{500}$ measured from mock clusters ($0<z<1.3$) confirm the picture shown by the observed sample. In fact, as for the observed clusters, all the simulation-based values of $N_{500}$ are lower than those predicted by the low-$z$ best-fitting relation of \citet{Lin-2004}, unlike those
measured by Lin et al. for the sample of \citet{De-Propris-1999} (Figs. \ref{fig:Figure5} \& \ref{fig:Figure8}, Table \ref{tab:Table3}). \\

\begin{figure}
\begin{center}
\includegraphics[width=0.5\textwidth,angle=0,clip]{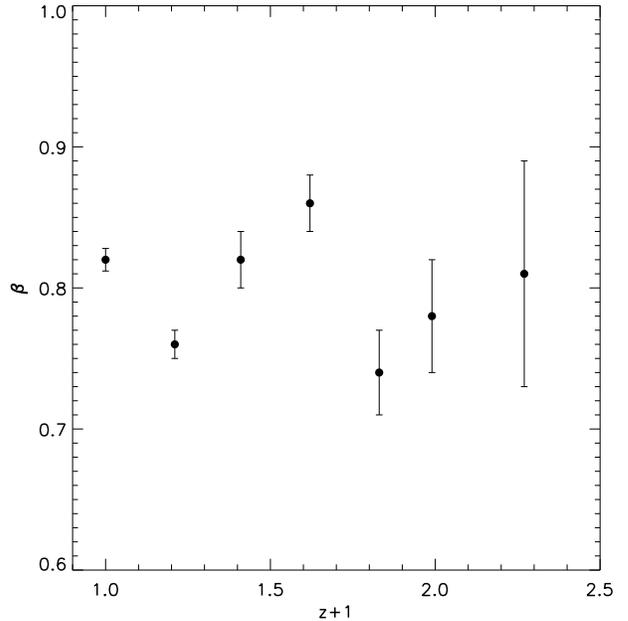} 
\caption{Halo HOD power-law index, as measured from mock clusters in each snapshot, against ($z$+1). Note all the clusters with 
$M_{200}>10^{14}\ {\rm M_{\odot}}$ in each snapshot are used to carry out this measure.}
\label{fig:Figure11}
\end{center}
\end{figure}

When the evolution of $\beta$ with $z$, as determined from mock cluster samples (Fig. \ref{fig:Figure11}), is studied, the values obtained 
seem to reveal no significant evolution with $z$ (as also indicated by a Spearmann's rank correlation 
test, which gives a correlation coefficient of $r=-0.2$ with a significance of its deviation from zero $p=0.6$). The value measured at $z=0$ ($\beta=0.820\pm0.008$) 
is consistent with the ones of \citet{Lin-2004}, \citet{Muzzin-2007b} and \citet{Popesso-2007}. 
Despite this, as noted in 
the introduction, many studies, e.g. \citet{Magliocchetti-2003, Zehavi-2004,
Rozo-2007} and \citet{Poggianti-2010}, measure discordant $\beta$ values ranging from $\beta=0.55\pm 0.043$ \citep{Marinoni-2002} to $\beta=1.1\pm 0.09$ \citep{Kochanek-2003}.
Particularly high are the values found by \citet{Abbas-2010}, who, depending on the magnitude depth used and the redshift ($0.1<z<1.3$), found 
$0.99\pm 0.10<\beta<1.66\pm 0.18$. However, the majority of the studies in the literature agree with $\beta$ being inconsistent with unity (this might not be the case for some
other simulation-based studies, but a meaningful comparison with these studies is particularly hard).\\ 

In order to compare the results obtained for cluster galaxies with dark matter, the predictions of the $c_{\rm dm}(z,M)$ relation obtained by \citet{Gao-2008} within the MS can be used as a reference.
Gao and collaborators presented in their study best-fitting $c_{\rm dm}(M_{200})$ relations at $z=0,0.5,1,2$ and 3. To carry out this comparison, the values of $c_{\rm dm}$ predicted by their model
are calculated using the values of $\overline{M}_{200}$ at $z=0,0.62$ and 0.99 for our MS mock clusters  (Table \ref{tab:Table3}, Fig. \ref{fig:Figure10}) and at $z\sim 1$ for our observed cluster sample ($c_{\rm dm}=3.55$).
This makes it possible to calculate $\Delta \log c=\log(c_{\rm g}/c_{\rm dm})$ \citep{Zheng-Weinberg-2007} at each of the chosen $z$ (Table \ref{tab:Table3}) and to compare galaxy spatial distribution
with the one of dark matter (a positive value of $\Delta \log c$ would mean galaxies are more clustered than dark matter and vice versa for negative values). This test shows that
$\Delta \log c$ decreases with redshift, reaching negative ($\Delta \log c=-0.1$, observed sample) or null ($\Delta \log c=0$, mock clusters) values at $z\sim 1$. At this $z$, the value
of $\Delta \log c$  is $< 5$ per cent of its value at $z=0$.\\

\subsubsection{Simulations: The Effect of Mass Selection}
\label{subsubsec:subsubsec6.2.3}
The results for $N_{500}$ and $c_{\rm g}$ shown in the previous sections are acquired using mock samples which have an increasing mean cluster mass with cosmic time (see Table \ref{tab:Table3}). As a result it is possible that this mass evolution affects the evolution of the HOD with cosmic time. In order to investigate this we repeat our analyses on two further mock samples of 100 clusters per snapshot: (i) selected within a constant mass range over cosmic time, set by the high-z sample; (ii) selected to have the same average mass at $z=1$ and $z=0$. We discuss the results from these samples below. \\

(i) We repeat our analysis on 100 mock clusters per snapshot randomly selected within the mass range of our observed cluster sample ($0.4<M_{500}<6.2 \times 10^{14}\ {\rm M}_{\odot}$). The values of the average masses of these mock cluster samples are $M_{500}=1.58$, $1.45$, $1.34$, $1.21$, $1.12$, $1.06$ and $1.04 \times 10^{14}\ {\rm M}_{\odot}$ at $z=0$, $0.21$, $0.41$, $0.62$, $0.83$, $0.99$ and $1.27$, respectively. When using these new mock clusters, we find very similar results to those described in Section \ref{subsubsec:subsubsec6.2.2} for the 100 most massive mock clusters at each snapshot. In particular we find $N_{500}/N_{\rm fit, local}$ is always $<1$ and for the 7 previously mentioned snapshots we find: $N_{500}/N_{\rm fit, local}=0.39\pm0.02$, $0.42\pm0.02$, $0.45\pm0.02$, $0.49\pm0.03$, $0.35\pm0.03$, $0.37\pm0.03$ and $0.53\pm0.03$ respectively.\\
We also recalculate the average concentration parameter using the $z=0$ and $z=1.27$ snapshots and again reproduce similar results within the errors: $c_{\rm g}=4.7^{+0.97}_{-0.81}$ at $z=1.27$ and
$c_{\rm g}=8.7^{+1.6}_{-1.3}$ at $z=0$. \\

(ii) As an additional test, we select mock clusters in the two redshift snapshots corresponding to $z=0$ and $z= 1.0$. At both these redshifts we generate a sample of 100 mock clusters, selected at random from a Gaussian centered at $M_{500}=2.1\times 10^{14} {\rm M}_{\odot}$, with a width given by the dispersion of our observed high-redshift clusters. Recomputing values in the same way as described in Section \ref{subsubsec:subsubsec6.2.2}, we find ${\langle N_{500} \rangle}_{z=1} = 28\pm1$ at $z=1$ and ${\langle N_{500} \rangle}_{z=0} = 36\pm1$ at  $z=0$. Normalising to the low-$z$ relation of \citet{Lin-2004} as before (see Fig. \ref{fig:Figure8}), we find again values less than 1: $N_{500}/N_{\rm fit, local}=0.62\pm0.02$ and  $N_{500}/N_{\rm fit, local}=0.5\pm0.02$ at $z=0$ and $z=1$ respectively. \\
For the concentration parameter we get $c_{\rm g}=7.5^{+0.96}_{-0.85}$ and $c_{\rm g}=3.7^{+0.5}_{-0.44}$ at $z=0$ and $z=1$, respectively. These results lie no more than $1-2 \sigma$ from the values at the corresponding redshift snapshots shown in Fig. \ref{fig:Figure10}, while the larger errors reflect the increase in Poisson noise due to the smaller number of galaxies in less rich clusters, which are more numerous in samples characterised by lower average masses.\\

The results obtained here for $N_{500}$ and $c_{\rm g}$ are both consistent with the simulation-based measurements in Fig. \ref{fig:Figure8} and Fig. \ref{fig:Figure10} respectively, which were based on mock samples whose average mass changes with epoch. They demonstrate the robustness of our claim that high-$z$ mock clusters agree with observations in hosting smaller numbers of galaxies compared to previous results and the claim of a significant evolutionary trend for $c_{\rm g}$. This conclusion is supported by other evidence in the literature which shows that the dependence of the concentration parameter on mass is weak for haloes more massive than $10^{14} {\rm M}_{\odot}$, e.g. \citet{Gao-2008}.\\ 

Finally, we note that the HOD power-law index $\beta$ evolutionary analysis shown in Fig. \ref{fig:Figure11} uses all clusters above $M_{200} = 10^{14} {\rm  M}_{\odot}$, in order to maximise the mass range and increase the signal given the large intrinsic scatter of the $N-M$ relation, with the consequence that the average mass in any of the 7 redshift snapshots varies by less than $30$ per cent from the overall average across all snapshots.

\begin{figure}
\begin{center}
\includegraphics[width=0.45\textwidth,angle=0,clip]{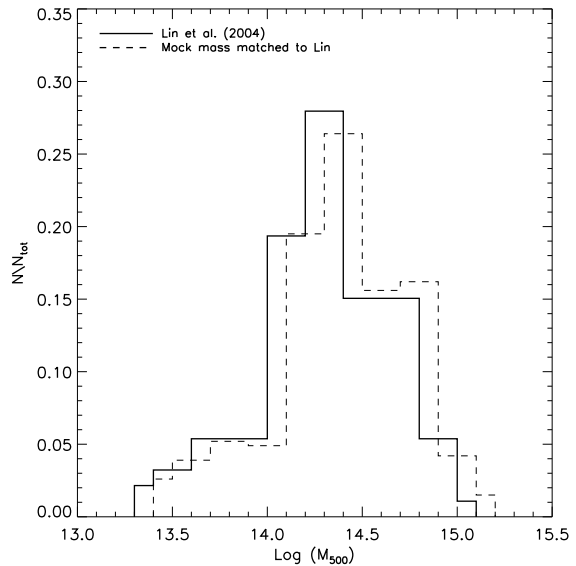}
\caption{Mass distributions of \citet{Lin-2004} cluster sample (full line) and its mass matched counterpart in the mocks (dashed line, offset of 0.1 {\rm dex} toward higher masses for clarity). See text for a description of the latter sample.}
\label{fig:Figure12}
\end{center}
\end{figure}

\subsubsection{Simulations: A Direct Comparison with Lin et al. at $z=0$}
\label{subsubsec:subsubsec6.2.4}
The mock sample at $z=0$ formed by the most massive haloes has a mass range ($5.2<M_{500}<26.0\ \times10^{14}\ {\rm M}_{\odot}$) which only partially overlaps with that of \citet{Lin-2004} ($0.2<M_{500}<12.6 \times 10^{14}\ {\rm M}_{\odot}$), resulting in a lack of haloes in the mocks less massive than $\sim 5 \times 10^{14}\ {\rm M}_{\odot}$. To investigate whether this disparity in the mass range is related to the large difference we find between our $z=0$ mock value of $c_{\rm g}$ and that of \citet{Lin-2004}, we generate 1000 clusters selected randomly from Lin's accumulated mass distribution and associate each of them with the $z=0$ mock cluster closest in mass. This produces a mock cluster sample closely matched to \citet{Lin-2004}; see Figure \ref{fig:Figure12}.
When using the new mock sample described we obtain $c_{\rm g}=8.0^{+0.32}_{-0.31}$, which is again consistent with the previous results for the low-z mocks and significantly different from the value ($c_{\rm g}=2.9^{+0.21}_{-0.22}$) of \citet{Lin-2004}.
 
\section{Discussion}
\label{sec:section7}
From our study of the evolution of ${\it K}^{\ast}$, shown in Fig. \ref{fig:Figure4}, it is evident that the SAMs reproduce the $K^{\ast}$ values of our high redshift clusters reasonably well. 
The data points of \citet{Cirasuolo-2007} and the parameterized fit of the evolution of  $K^{\ast}$ from \citet{Cirasuolo-2010} for 
the field LF are also plotted in this Figure. It can be seen there is a small difference of, on average, $\Delta {\it K}^{\ast} \sim 0.5$ at $z>0.8$ between the cluster environment (brighter ${\it K}^{\ast}$) and the field. Furthermore, \citet{Balogh-2001} found in their study of the {\it J}-band LF a difference of similar magnitude between cluster and field of $\Delta {\it J}^{\ast}\sim0.5$. The results of 
Cirasuolo et al., are from a sample of 50,000 galaxies with photometric redshifts $0.25 \leq z \leq 3.25$ from the UKIDSS UDS in which they compare the {\it K}-band LF with the predictions of SAMs (including \citealp{Bower-2006}) finding broad consistency - even though some models still tend to under predict the number of field galaxies at high $z$. However, we point out that the \citet{Bower-2006} SAM does a better job of predicting the number of high-$z$ field galaxies than other SAMs. Although our study is restricted to the analysis of $K^{\ast}$, it is clear that our results indicate 
that the current SAMs can predict the luminosities of $L^{\ast}$ galaxies both in the field and rich clusters, where substantially more merging has taken place, reasonably well. On the other hand some authors  (e.g., \citealp{Cimatti-2006, Pozzetti-2010}) have pointed out that the massive end of the galaxy mass function does not appear to evolve significantly since $z\sim 1$ and questioned whether downsizing is compatible with the predictions of SAMs.
As mentioned in the introduction, the situation is potentially even worse for BCGs at the centre of rich clusters. In fact, Collins et al (2009) and Stott et al (2010) showed that the observed 
stellar mass of BCGs in high-mass X-ray clusters does not evolve significantly since $z\simeq1.5$, in contrast with the  SAMs which predict that BCGs at this redshift should only have $20-30$ per cent of the  
stellar mass of their local counterparts.

These results, taken together with the small difference between observed field and cluster $K^{\ast}$ values, now confirmed here, (see also \citealp{Cole-2001, De-Propris-2009}) and together with other evidence found in the literature that the luminous ($L>L^{\ast}$) part of the LF weakly depends on the environment (e.g., \citealp{Cimatti-2006, De-Filippis-2011}), indicates that the hierarchical galaxy formation scenario
works well for galaxies up to a mass $M\sim 10^{11}\ {\rm {M_{\odot}}}$ (about the mass of an $L^{\ast}$ galaxy). \\
 
Our halo HOD power-law indexes ($\beta \simeq 0.8$) for mock clusters are in agreement with several studies in the literature, e.g. \citet{Lin-2004}, \citet{Muzzin-2007b} 
and \citet{Popesso-2007}. However, the picture is complicated, since several
other studies such as \citet{Kochanek-2003} and \citet{Poggianti-2010}, found values of $\beta$ greater than or consistent with 1. As 
emphasized by Popesso and collaborators and Lin et al., a value $\beta<1$ is expected from the theoretical point of view.
This is because, despite the fact that the hierarchical structure formation models (e.g. \citealp{De-Lucia-Kauffmann-2004, Gao-2004}) predict a universal 
mass distribution of sub-haloes (independent of the parent halo's mass and with the consequence that the number of sub-haloes is 
proportional to halo mass), the introduction of baryons produces a decreasing number of galaxies per unit mass (e.g., \citealp{Berlind-2003}) 
and an increasing $M/L$ ratio (e.g., \citealp{Kauffmann-1999}) for higher halo masses. Several processes may be responsible for this
behavior, such as an increased merger rate \citep{White-2001} or an increasing galaxy destruction rate 
\citep{Lin-2004} or decreasing star formation and gas cooling efficiencies (e.g., \citealp{Berlind-2003}). However, as pointed out by 
\citet{Popesso-2007}, each of these processes should leave their mark in the properties of galaxy clusters. \citet{Lin-2004} split their 
sample into a high and low-mass sub-samples and found both an excess and a lack of galaxies in the faint and 
the bright parts of the LF respectively for the low-mass cluster sub-sample, although they concluded that none of the studied processes 
(tidal stripping, ram pressure stripping, galaxy harassment, variations in star formation efficiency, stellar aging, differences in 
galaxy radial distribution, dynamical friction and galaxy mergers) can by itself satisfactorily explain these differences. On the other 
hand, \citet{Popesso-2007} found a universal LF with no dependence on the cluster mass and wave-band. For this reason, they proposed the 
only way to have $\beta<1$ leaving the properties of galaxy clusters unchanged is if the sub-haloes mass distribution is not universal.
It is beyond the scope of the present study to address which process may cause 
$\beta<1$ or whether the sub-haloes mass distribution is universal and we simply point out that our analysis on mock clusters shows no 
significant evolution of $\beta$ with $z$ ($\beta\sim 0.8$ out to $z\sim1.3$) and that whatever 
this process might be, it is already in place by $z\sim 1$. \\

Our observational-based results indicate that low-$z$ clusters are richer than their high-$z$ counterparts when comparing  clusters of the same mass. Using
the best fit relation to Eq. \ref{eq:eq12} and the masses in Table \ref{tab:Table3}, $N_{500}$  increase by a factor of $\sim 1.6$ from $z\sim 1$ to $z\sim 0.6$ and by a factor of $\sim 2.6$ from 
$z\sim 0.6$ to $z=0$. This growth rate of the number of galaxies is very similar to the one found by \citet{Abbas-2010} over a similar redshift interval.
 
Using mock clusters, \citet{Poggianti-2010} also found little evolution in the $\beta$ value for their samples, although pointed out that values can depend on whether masses of mock clusters are derived from their intrinsic halo mass ($\beta=1.00\pm0.04$) or using their velocity dispersion mass proxy ($\beta=0.77\pm0.03$). It is not possible to match our X-ray measurements to the simulations in the same way at the present time but clearly further work is necessary to compare different mass estimators. However, we point out that we obtain an average value of $\beta<1$ for our mock clusters despite using the halo mass values provided by the simulation. \\
   
Focusing the attention on the galaxy concentration parameter (Table \ref{tab:Table3} and Figs. \ref{fig:Figure6} \& \ref{fig:Figure10}), the present analysis leads to a picture which is difficult 
to fully understand, partly due to the diversity of the $c_{\rm g}$ found in the literature \citep{Carlberg-1997, van-der-Marel-2000, Biviano-2003, 
Katgert-2004, Lin-2004, Rines-2006, Muzzin-2007a, Biviano-2010}, whose large scatter is probably due to the heterogeneity of the analysis methods, which are often based on different radial distances and magnitude limits. Another possible cause of this large scatter could be the cluster-to-cluster variation, which can influence the headline value of $c_{\rm g}$ when simply averaging over individual measurements (using individual surface density profiles), as opposed to the preferred method of fitting the stacked density profile. However, we note that the most accurate $c_{\rm g}$ estimate on intermediate $z$ scale ($c_{\rm g}=4.13\pm0.57$), carried out by \citet{Muzzin-2007a} and based on redshift measurements, agrees well with our observations at $z=1$.
The values of $c_{\rm g}$ measured for our mock clusters are seen to increase with decreasing 
redshift, giving rise to a significant correlation with a Spearman rank correlation coefficient of $-0.96$. In particular, those obtained for high-$z$ mock clusters ($c_{\rm g}=3.85_{-0.4}^{+0.45}, 3.70_{-0.44}^{+0.5}$ and 
$3.77_{-0.51}^{+0.58}$ at $z=$0.83, 0.99 and 1.27, respectively) are found consistent within $1\sigma$ with the value of 
$c_{\rm g}=2.8_{-0.8}^{+1}$ found for our observed sample at $z\sim 1$. The large errors characterising the observed values found in the literature, do not allow us to address this issue in a more quantitative way. Having said that, when the value obtained for mock clusters at $z=0$ ($c_{\rm g}=6.30_{-0.36}^{+0.39}$) is compared with the observed one of \citet{Lin-2004} ($c_{\rm g}=2.90_{-0.22}^{+0.21}$) at $z\sim 0.06$, the discrepancy appears large and significant, although the models agree with other low-$z$ observation-based estimates of $c_{\rm g}$. If real, this would be an indication that SAMs have problems in predicting the evolution of cluster galaxies observed from $z\sim 1$ to $z=0$. \\

A potentially very important difficulty which arises in the simulations is the problem of the {\it orphan} galaxies (Shaun Cole, private communication). These are 
galaxies which have lost their dark matter haloes and whose tracks in the simulations are 
traced by following the particle that was the one most bound in its  halo prior to disruption. In the \citet{Bower-2006} SAM,
the merging of satellites onto the central galaxy is treated  using a dynamical friction calculation that, instead of using the true orbit
of the galaxy, uses random orbital parameters that are only statistically consistent with N-body simulations. This could lead
to biasing the concentration parameter towards higher values. In fact, in reality, the satellites most likely to merge are the ones on orbits
closest to the centre, whereas random selection in the SAM of Bower et al. removes satellites randomly at all radii. Hence, the
remaining satellites may be too concentrated. While there is no reason to expect that the current treatment of satellite galaxy mergers should give rise to the entire trend seen in Fig. \ref{fig:Figure10}, the possibility cannot be ruled out. This issue will only be decided by carrying out a more accurate
treatment of the merging and orbits of satellite galaxies, which has already been developed, but not yet applied to the MS. However, it is worth pointing out that the problems of calculating the radial distribution of satellite galaxies in the \citet{Bower-2006} model are not expected to lead to significant errors in 
the numbers of satellite galaxies contributing to the HOD.\\

The growth of mock $c_{\rm g}$ with decreasing $z$ seen in the present analysis is not unexpected. What is surprising, instead, is the rate of 
this growth. Investigating dark matter haloes within simulations, 
studies like \citet{Neto-2007, Gao-2008, Duffy-2008} and \citet{Munoz-2011} addressed the $c_{\rm dm}(z,M)$ relation in details finding 
$c_{\rm dm}$ being anti-correlated with mass and $z$. In particular, the anti-correlation between $c_{\rm dm}$ and $M$ weakens at higher $z$ and 
evidence of dark matter haloes having a constant $c_{\rm dm}\sim 3.5-4$ at $z>1$ have been found by \citet{Gao-2008}. \\

Comparing the mock-cluster values of $c_{\rm g}$ measured here with the ones of the dark matter haloes 
predicted by theoretical studies \citep{Gao-2008} at different $z$ (see Section \ref{sec:section6}), the results seem to show that galaxies have a similar concentration to 
dark matter at $z\sim 1$ but subsequently they become more concentrated with decreasing redshift. This may be due to the effect of non-gravitational or gravitational processes, such as: gas cooling, AGN feedback, dynamical friction, merging and tidal stripping, which have significantly modified the galaxy distribution within the dark 
matter halo over the last $\sim 9$ {\rm Gyr}. This growth is dominated by the continuously refined physics of the SAMs (e.g., \citealp{Duffy-2010}), and given this sensitivity, a glance at the data in Fig. \ref{fig:Figure10} suggests the need for a thorough and homogeneous study of $c_{\rm g}$ in observed galaxy clusters and groups over a wide range in redshift.\\

\section{Conclusions}
\label{sec:section8}
We study the evolution of cluster galaxies with $z$ using {\it K}-band photometry of a cluster sample made of 15 of the highest-$z$ ($0.8<z<1.5$) 
X-ray clusters observed so far. In particular, we investigate the ${\it K}^{\ast}$ Hubble diagram out to $z\sim 1.3$ in comparison with SSP 
(no evolution in  absolute {\it K}-band magnitude and passive evolution) and SAM evolutionary models.  This allows us to explore the process of galaxy formation. 
We also study the HOD of this sample, by investigating 
$\langle N \rangle$, as a function of $M$ and $z$, and $c_{\rm g}$ with $z$ in comparison with $c_{\rm dm}$. All our analysis is carried out in 
strict comparison with simulations, through the use of mock cluster samples taken from MS at $0<z<1.3$ and populated with galaxies by means of 
the SAM by \citet{Bower-2006}. In addition to being an important test for SAMs, this allows us to investigate in details the evolution of $\beta$, 
$\langle N \rangle$ and $c_{\rm g}$ with $z$ and to study how the relationship between galaxies and dark matter  
spatial distributions within haloes changes with cosmic time. This kind of study is fundamental for understanding the role non-gravitational processes, as part
of the physics of galaxy formation, take in influencing the way galaxies populate dark matter haloes. Our main conclusions are: 

\begin{itemize}

\item[(i)] The values of ${\it K}^{\ast}$ obtained show the existence of old, evolved and massive galaxies at $z>1.5$. Despite the evolution of ${\it K}^{\ast}$ is well reproduced
                by SSP passive evolutionary models ($z_{\rm f}>3$),  we can not disentangle between these models and the predictions of \citet{Bower-2006} SAM, which is seen, for 
                the first time, to reproduce well the evolution of ${\it K}^{\ast}$ also in the cluster environment. \\
  
\item[(ii)] When comparing the values of ${\it K}^{\ast}$ obtained for observed clusters with the ones of the field, there are no major differences. This fact, coupled with the marginal evolution
                of the massive end of the galaxy mass function seen observationally (e.g., \citealp{Cimatti-2006, Pozzetti-2010}), leads to questioning whether the role of the environment is
                negligible for the formation of galaxies more massive than $\sim10^{11}\ {\rm {M_{\odot}}}$ (about the mass of a $L^{\ast}$ galaxy).\\
                   
\item[(iii)] By investigating $N_{500}$ in real and mock clusters, this study shows that high-$z$ clusters are poorer than those at low $z$.
                 Unlike \citet{Lin-2006}, we find significant trends of $N_{500}$ with both $z$ and cluster mass: 
                 $N(M,z)=(53\pm1)(1+z)^{-0.61^{+0.18}_{-0.20}} (M/10^{14.3})^{0.86\pm0.05}$. \\ 
                 
\item[(iv)] Using mock clusters, the slope $\beta$ of the $N_{500}-M_{500}$ relation is found to be significantly lower than one at all $z$ (out to $z\sim 1.3$), showing no 
                 significant signs of evolution. This means more massive clusters are characterized by a lower galaxy number per unit mass compared to lower mass systems already 
                 at high $z$. Because of this, the local value of $\beta<1$ can not be explained as being due to local clusters found to be richer than those at high-$z$. \\      
   
\item[(v)]  Although our results here and those from the literature seem to indicate a decreasing trend of $c_{\rm g}$ with $z$, overall the data and SAM-based predictions are very uncertain and the current situation emphasizes the compelling need of a
                 systematic and homogeneous study of the galaxy concentration parameter in clusters over low and intermediate redshifts.\\

\item[(vi)]  When comparing our mock-cluster values of $c_{\rm g}$ (3.7-6.3) with those of  $c_{\rm dm}$ (3.6-3.8), starting from similar concentrations at $z\sim 1$, galaxies seem to become more 
                  concentrated than dark matter as $z$ approaches 0. This may be due to gravitational and/or non-gravitational processes significantly modifying the distribution of galaxies within dark matter
                  haloes over the last $\sim 9$ {\rm Gyr}. However, the problem of the {\it orphan galaxies} in the SAMs prevents us concluding that this is a real physical trend.

\end{itemize}

\subsection*{Acknowledgments}
The authors thank the referee Cedric Lacey for a report which significantly improved the paper. We also thank Shaun Cole for the useful discussion about the SAM models and Adam Muzzin for kindly 
providing his data.
DC expresses his gratitude to Ivan Baldry and Maurizio Paolillo for the valuable discussions and support.  DC also thanks Gerard Lemson for the helpful suggestions with regards to the MS database. 
CAC and JPS acknowledge financial support from STFC. MH acknowledges financial support from the Leverhulme Trust. This work is based in part on data collected at the Subaru Telescope, which is operated by the National Astronomy Observatory of Japan and the XMM-Newton, an ESA science mission funded by contributions from ESA member states and from NASA.\\
IRAF is distributed by the National Optical Astronomy Observatories, which are operated by the Association of Universities for Research in Astronomy, Inc., under cooperative agreement with the
National Science Foundation.

\bibliographystyle{mn2e}

\bibliography{Reference}

\begin{thebibliography}{101}
\expandafter\ifx\csname natexlab\endcsname\relax\def\natexlab#1{#1}\fi

\bibitem[{{Abbas} {et~al}\mbox{.}(2010){Abbas} {et~al.}}]{Abbas-2010}
{Abbas} U., {et~al.}, 2010, \mnras, 406, 1306

\bibitem[{{Andreon}(2006)}]{Andreon-2006}
{Andreon} S., 2006, \aap, 448, 447

\bibitem[{{Andreon} {et~al}\mbox{.}(2008){Andreon}, {De Propris}, {Puddu},
  {Giordano}, \& {Quintana}}]{Andreon-2008}
{Andreon} S., {De Propris} R., {Puddu} E., {Giordano} L., {Quintana} H., 2008,
  \mnras, 383, 102

\bibitem[{{Balestra} {et~al}\mbox{.}(2007){Balestra}, {Tozzi}, {Ettori},
  {Rosati}, {Borgani}, {Mainieri}, {Norman}, \& {Viola}}]{Balestra-2007}
{Balestra} I., {Tozzi} P., {Ettori} S., {Rosati} P., {Borgani} S., {Mainieri}
  V., {Norman} C., {Viola} M., 2007, \aap, 462, 429

\bibitem[{{Balogh} {et~al}\mbox{.}(2001){Balogh}, {Christlein}, {Zabludoff}, \&
  {Zaritsky}}]{Balogh-2001}
{Balogh} M.~L., {Christlein} D., {Zabludoff} A.~I., {Zaritsky} D., 2001, \apj,
  557, 117

\bibitem[{{Bartelmann}(1996)}]{Bartelmann-1996}
{Bartelmann} M., 1996, \aap, 313, 697

\bibitem[{{Benson} {et~al}\mbox{.}(2000){Benson}, {Baugh}, {Cole}, {Frenk}, \&
  {Lacey}}]{Benson-2000a}
{Benson} A.~J., {Baugh} C.~M., {Cole} S., {Frenk} C.~S., {Lacey} C.~G., 2000,
  \mnras, 316, 107

\bibitem[{{Berlind} \& {Weinberg}(2002)}]{Berlind-2002}
{Berlind} A.~A., {Weinberg} D.~H., 2002, \apj, 575, 587

\bibitem[{{Berlind} {et~al}\mbox{.}(2003){Berlind}, {Weinberg}, {Benson},
  {Baugh}, {Cole}, {Dav{\'e}}, {Frenk}, {Jenkins}, {Katz}, \&
  {Lacey}}]{Berlind-2003}
{Berlind} A.~A. {et~al.}, 2003, \apj, 593, 1

\bibitem[{{Biviano} \& {Girardi}(2003)}]{Biviano-2003}
{Biviano} A., {Girardi} M., 2003, \apj, 585, 205

\bibitem[{{Biviano} \& {Poggianti}(2010)}]{Biviano-2010}
{Biviano} A., {Poggianti} B., 2010, in American Institute of Physics Conference
  Series, Vol. 1241, American Institute of Physics Conference Series,
  {J.-M.~Alimi \& A.~Fu{\"o}zfa}, ed., pp. 192--199

\bibitem[{{Bower} {et~al}\mbox{.}(2006){Bower}, {Benson}, {Malbon}, {Helly},
  {Frenk}, {Baugh}, {Cole}, \& {Lacey}}]{Bower-2006}
{Bower} R.~G., {Benson} A.~J., {Malbon} R., {Helly} J.~C., {Frenk} C.~S.,
  {Baugh} C.~M., {Cole} S., {Lacey} C.~G., 2006, \mnras, 370, 645

\bibitem[{{Branchesi} {et~al}\mbox{.}(2007){Branchesi}, {Gioia}, {Fanti}, \&
  {Fanti}}]{Branchesi-2007}
{Branchesi} M., {Gioia} I.~M., {Fanti} C., {Fanti} R., 2007, \aap, 472, 739

\bibitem[{{Bremer} {et~al}\mbox{.}(2006){Bremer}, {Valtchanov}, {Willis},
  {Altieri}, {Andreon}, {Duc}, {Fang}, {Jean}, {Lonsdale}, {Pacaud}, {Pierre},
  {Shupe}, {Surace}, \& {Waddington}}]{Bremer-2006}
{Bremer} M.~N. {et~al.}, 2006, \mnras, 371, 1427

\bibitem[{{Bruzual} \& {Charlot}(2003)}]{Bruzual-2003}
{Bruzual} G., {Charlot} S., 2003, \mnras, 344, 1000

\bibitem[{{Buote} {et~al}\mbox{.}(2007){Buote}, {Gastaldello}, {Humphrey},
  {Zappacosta}, {Bullock}, {Brighenti}, \& {Mathews}}]{Buote-2007}
{Buote} D.~A., {Gastaldello} F., {Humphrey} P.~J., {Zappacosta} L., {Bullock}
  J.~S., {Brighenti} F., {Mathews} W.~G., 2007, \apj, 664, 123

\bibitem[{{Capozzi}, {Collins} \& {Stott}(2010){Capozzi}, {Collins}, \&
  {Stott}}]{Capozzi-2010}
{Capozzi} D., {Collins} C.~A., {Stott} J.~P., 2010, \mnras, 403, 1274

\bibitem[{{Carlberg} {et~al}\mbox{.}(1997){Carlberg}, {Yee}, {Ellingson},
  {Morris}, {Abraham}, {Gravel}, {Pritchet}, {Smecker-Hane}, {Hartwick},
  {Hesser}, {Hutchings}, \& {Oke}}]{Carlberg-1997}
{Carlberg} R.~G. {et~al.}, 1997, \apjl, 485, L13+

\bibitem[{{Christlein} \& {Zabludoff}(2003)}]{Christlein-2003}
{Christlein} D., {Zabludoff} A.~I., 2003, \apj, 591, 764

\bibitem[{{Cimatti}, {Daddi} \& {Renzini}(2006){Cimatti}, {Daddi}, \&
  {Renzini}}]{Cimatti-2006}
{Cimatti} A., {Daddi} E., {Renzini} A., 2006, \aap, 453, L29

\bibitem[{{Cirasuolo} {et~al}\mbox{.}(2010){Cirasuolo}, {McLure}, {Dunlop},
  {Almaini}, {Foucaud}, \& {Simpson}}]{Cirasuolo-2010}
{Cirasuolo} M., {McLure} R.~J., {Dunlop} J.~S., {Almaini} O., {Foucaud} S.,
  {Simpson} C., 2010, \mnras, 401, 1166

\bibitem[{{Cirasuolo} {et~al}\mbox{.}(2007){Cirasuolo}, {McLure}, {Dunlop},
  {Almaini}, {Foucaud}, {Smail}, {Sekiguchi}, {Simpson}, {Eales}, {Dye},
  {Watson}, {Page}, \& {Hirst}}]{Cirasuolo-2007}
{Cirasuolo} M. {et~al.}, 2007, \mnras, 380, 585

\bibitem[{{Cole} {et~al}\mbox{.}(2001){Cole} {et~al.}}]{Cole-2001}
{Cole} S., {et~al.}, 2001, \mnras, 326, 255

\bibitem[{{Colless} {et~al}\mbox{.}(2001){Colless} {et~al.}}]{Colless-2001}
{Colless} M., {et~al.}, 2001, \mnras, 328, 1039

\bibitem[{{Collins} {et~al}\mbox{.}(2009){Collins}, {Stott}, {Hilton}, {Kay},
  {Stanford}, {Davidson}, {Hosmer}, {Hoyle}, {Liddle}, {Lloyd-Davies}, {Mann},
  {Mehrtens}, {Miller}, {Nichol}, {Romer}, {Sahl{\'e}n}, {Viana}, \&
  {West}}]{Collins-2009}
{Collins} C.~A. {et~al.}, 2009, \nat, 458, 603

\bibitem[{{Collister} \& {Lahav}(2005)}]{Collister-Lahav-2005}
{Collister} A.~A., {Lahav} O., 2005, \mnras, 361, 415

\bibitem[{{Comerford} \& {Natarajan}(2007)}]{Comerford-2007}
{Comerford} J.~M., {Natarajan} P., 2007, \mnras, 379, 190

\bibitem[{{Cowie} {et~al}\mbox{.}(1996){Cowie}, {Songaila}, {Hu}, \&
  {Cohen}}]{Cowie-1996}
{Cowie} L.~L., {Songaila} A., {Hu} E.~M., {Cohen} J.~G., 1996, \aj, 112, 839

\bibitem[{{De Filippis} {et~al}\mbox{.}(2011){De Filippis}, {Paolillo},
  {Longo}, {La Barbera}, {de Carvalho}, \& {Gal}}]{De-Filippis-2011}
{De Filippis} E., {Paolillo} M., {Longo} G., {La Barbera} F., {de Carvalho}
  R.~R., {Gal} R., 2011, \mnras, 414, 2771

\bibitem[{{De Lucia} \& {Blaizot}(2007)}]{De-Lucia-Blaizot-2007}
{De Lucia} G., {Blaizot} J., 2007, \mnras, 375, 2

\bibitem[{{De Lucia} {et~al}\mbox{.}(2004){De Lucia}, {Kauffmann}, {Springel},
  {White}, {Lanzoni}, {Stoehr}, {Tormen}, \&
  {Yoshida}}]{De-Lucia-Kauffmann-2004}
{De Lucia} G., {Kauffmann} G., {Springel} V., {White} S.~D.~M., {Lanzoni} B.,
  {Stoehr} F., {Tormen} G., {Yoshida} N., 2004, \mnras, 348, 333

\bibitem[{{De Lucia} {et~al}\mbox{.}(2006){De Lucia}, {Springel}, {White},
  {Croton}, \& {Kauffmann}}]{De-Lucia-2006}
{De Lucia} G., {Springel} V., {White} S.~D.~M., {Croton} D., {Kauffmann} G.,
  2006, \mnras, 366, 499

\bibitem[{{De Lucia} {et~al}\mbox{.}(2007){De Lucia} {et~al.}}]{De-Lucia-2007}
{De Lucia} G., {et~al.}, 2007, \mnras, 374, 809

\bibitem[{{De Propris} \& {Christlein}(2009)}]{De-Propris-2009}
{De Propris} R., {Christlein} D., 2009, Astronomische Nachrichten, 330, 943

\bibitem[{{De Propris} {et~al}\mbox{.}(1998){De Propris}, {Eisenhardt},
  {Stanford}, \& {Dickinson}}]{De-Propris-1998}
{De Propris} R., {Eisenhardt} P.~R., {Stanford} S.~A., {Dickinson} M., 1998,
  \apjl, 503, L45+

\bibitem[{{De Propris} {et~al}\mbox{.}(1999){De Propris}, {Stanford},
  {Eisenhardt}, {Dickinson}, \& {Elston}}]{De-Propris-1999}
{De Propris} R., {Stanford} S.~A., {Eisenhardt} P.~R., {Dickinson} M., {Elston}
  R., 1999, \aj, 118, 719

\bibitem[{{Duffy} {et~al}\mbox{.}(2008){Duffy}, {Schaye}, {Kay}, \& {Dalla
  Vecchia}}]{Duffy-2008}
{Duffy} A.~R., {Schaye} J., {Kay} S.~T., {Dalla Vecchia} C., 2008, \mnras, 390,
  L64

\bibitem[{{Duffy} {et~al}\mbox{.}(2010){Duffy}, {Schaye}, {Kay}, {Dalla
  Vecchia}, {Battye}, \& {Booth}}]{Duffy-2010}
{Duffy} A.~R., {Schaye} J., {Kay} S.~T., {Dalla Vecchia} C., {Battye} R.~A.,
  {Booth} C.~M., 2010, \mnras, 405, 2161

\bibitem[{{Ellis} \& {Jones}(2004)}]{Ellis-2004}
{Ellis} S.~C., {Jones} L.~R., 2004, \mnras, 348, 165

\bibitem[{{Gao} {et~al}\mbox{.}(2008){Gao}, {Navarro}, {Cole}, {Frenk},
  {White}, {Springel}, {Jenkins}, \& {Neto}}]{Gao-2008}
{Gao} L., {Navarro} J.~F., {Cole} S., {Frenk} C.~S., {White} S.~D.~M.,
  {Springel} V., {Jenkins} A., {Neto} A.~F., 2008, \mnras, 387, 536

\bibitem[{{Gao} {et~al}\mbox{.}(2004){Gao}, {White}, {Jenkins}, {Stoehr}, \&
  {Springel}}]{Gao-2004}
{Gao} L., {White} S.~D.~M., {Jenkins} A., {Stoehr} F., {Springel} V., 2004,
  \mnras, 355, 819

\bibitem[{{Hashimoto} {et~al}\mbox{.}(2004){Hashimoto}, {Barcons},
  {B{\"o}hringer}, {Fabian}, {Hasinger}, {Mainieri}, \&
  {Brunner}}]{Hashimoto-2004}
{Hashimoto} Y., {Barcons} X., {B{\"o}hringer} H., {Fabian} A.~C., {Hasinger}
  G., {Mainieri} V., {Brunner} H., 2004, \aap, 417, 819

\bibitem[{{Hicks} {et~al}\mbox{.}(2008){Hicks}, {Ellingson}, {Bautz}, {Cain},
  {Gilbank}, {Gladders}, {Hoekstra}, {Yee}, \& {Garmire}}]{Hicks-2008}
{Hicks} A.~K. {et~al.}, 2008, \apj, 680, 1022

\bibitem[{{Hilton} {et~al}\mbox{.}(2010){Hilton} {et~al.}}]{Hilton-2010}
{Hilton} M., {et~al.}, 2010, \apj, 718, 133

\bibitem[{{Ho} {et~al}\mbox{.}(2009){Ho}, {Lin}, {Spergel}, \&
  {Hirata}}]{Ho-2009}
{Ho} S., {Lin} Y.-T., {Spergel} D., {Hirata} C.~M., 2009, \apj, 697, 1358

\bibitem[{{Hu} \& {Kravtsov}(2003)}]{Hu-2003}
{Hu} W., {Kravtsov} A.~V., 2003, \apj, 584, 702

\bibitem[{{Ichikawa} {et~al}\mbox{.}(2006){Ichikawa}, {Suzuki}, {Tokoku},
  {Uchimoto}, {Konishi}, {Yoshikawa}, {Yamada}, {Tanaka}, {Omata}, \&
  {Nishimura}}]{Ichikawa-2006}
{Ichikawa} T. {et~al.}, 2006, in Society of Photo-Optical Instrumentation
  Engineers (SPIE) Conference Series, Vol. 6269, Society of Photo-Optical
  Instrumentation Engineers (SPIE) Conference Series

\bibitem[{{Jing}(2000)}]{Jing-2000}
{Jing} Y.~P., 2000, \apj, 535, 30

\bibitem[{{Katgert}, {Biviano} \& {Mazure}(2004){Katgert}, {Biviano}, \&
  {Mazure}}]{Katgert-2004}
{Katgert} P., {Biviano} A., {Mazure} A., 2004, \apj, 600, 657

\bibitem[{{Kauffmann} {et~al}\mbox{.}(1999){Kauffmann}, {Colberg}, {Diaferio},
  \& {White}}]{Kauffmann-1999}
{Kauffmann} G., {Colberg} J.~M., {Diaferio} A., {White} S.~D.~M., 1999, \mnras,
  303, 188

\bibitem[{{Kochanek} {et~al}\mbox{.}(2003){Kochanek}, {White}, {Huchra},
  {Macri}, {Jarrett}, {Schneider}, \& {Mader}}]{Kochanek-2003}
{Kochanek} C.~S., {White} M., {Huchra} J., {Macri} L., {Jarrett} T.~H.,
  {Schneider} S.~E., {Mader} J., 2003, \apj, 585, 161

\bibitem[{{Kravtsov} {et~al}\mbox{.}(2004){Kravtsov}, {Berlind}, {Wechsler},
  {Klypin}, {Gottl{\"o}ber}, {Allgood}, \& {Primack}}]{Kravtsov-2004}
{Kravtsov} A.~V., {Berlind} A.~A., {Wechsler} R.~H., {Klypin} A.~A.,
  {Gottl{\"o}ber} S., {Allgood} B., {Primack} J.~R., 2004, \apj, 609, 35

\bibitem[{{Lamer} {et~al}\mbox{.}(2008){Lamer}, {Hoeft}, {Kohnert}, {Schwope},
  \& {Storm}}]{Lamer-2008}
{Lamer} G., {Hoeft} M., {Kohnert} J., {Schwope} A., {Storm} J., 2008, \aap,
  487, L33

\bibitem[{{Lawrence} {et~al}\mbox{.}(2007){Lawrence} {et~al.}}]{Lawrence-2007}
{Lawrence} A., {et~al.}, 2007, \mnras, 379, 1599

\bibitem[{{Leggett}(1992)}]{Leggett-1992}
{Leggett} S.~K., 1992, \apjs, 82, 351

\bibitem[{{Lemson} \& {Virgo Consortium}(2006)}]{Lemson-2006}
{Lemson} G., {Virgo Consortium} t., 2006, ArXiv Astrophysics e-prints: 0608019

\bibitem[{{Lin} {et~al}\mbox{.}(2006){Lin}, {Mohr}, {Gonzalez}, \&
  {Stanford}}]{Lin-2006}
{Lin} Y., {Mohr} J.~J., {Gonzalez} A.~H., {Stanford} S.~A., 2006, \apjl, 650,
  L99

\bibitem[{{Lin}, {Mohr} \& {Stanford}(2004){Lin}, {Mohr}, \&
  {Stanford}}]{Lin-2004}
{Lin} Y., {Mohr} J.~J., {Stanford} S.~A., 2004, \apj, 610, 745

\bibitem[{{Lubin}, {Mulchaey} \& {Postman}(2004){Lubin}, {Mulchaey}, \&
  {Postman}}]{Lubin-2004}
{Lubin} L.~M., {Mulchaey} J.~S., {Postman} M., 2004, \apjl, 601, L9

\bibitem[{{Magliocchetti} \& {Porciani}(2003)}]{Magliocchetti-2003}
{Magliocchetti} M., {Porciani} C., 2003, \mnras, 346, 186

\bibitem[{{Marinoni} \& {Hudson}(2002)}]{Marinoni-2002}
{Marinoni} C., {Hudson} M.~J., 2002, \apj, 569, 101

\bibitem[{{Martini}(2001)}]{Martini-2001}
{Martini} P., 2001, \aj, 121, 598

\bibitem[{{Maughan}(2007)}]{Maughan-2007}
{Maughan} B.~J., 2007, \apj, 668, 772

\bibitem[{{Maughan} {et~al}\mbox{.}(2004){Maughan}, {Jones}, {Ebeling}, \&
  {Scharf}}]{Maughan-2004}
{Maughan} B.~J., {Jones} L.~R., {Ebeling} H., {Scharf} C., 2004, \mnras, 351,
  1193

\bibitem[{{Maughan} {et~al}\mbox{.}(2006){Maughan}, {Jones}, {Ebeling}, \&
  {Scharf}}]{Maughan-2006}
{Maughan} B.~J., {Jones} L.~R., {Ebeling} H., {Scharf} C., 2006, \mnras, 365,
  509

\bibitem[{{McCracken} {et~al}\mbox{.}(2010){McCracken}
  {et~al.}}]{McCracken-2010}
{McCracken} H.~J., {et~al.}, 2010, \apj, 708, 202

\bibitem[{{Mu{\~n}oz-Cuartas} {et~al}\mbox{.}(2011){Mu{\~n}oz-Cuartas},
  {Macci{\`o}}, {Gottl{\"o}ber}, \& {Dutton}}]{Munoz-2011}
{Mu{\~n}oz-Cuartas} J.~C., {Macci{\`o}} A.~V., {Gottl{\"o}ber} S., {Dutton}
  A.~A., 2011, \mnras, 411, 584

\bibitem[{{Muzzin} {et~al}\mbox{.}(2007{\natexlab{a}}){Muzzin}, {Yee}, {Hall},
  {Ellingson}, \& {Lin}}]{Muzzin-2007a}
{Muzzin} A., {Yee} H.~K.~C., {Hall} P.~B., {Ellingson} E., {Lin} H.,
  2007{\natexlab{a}}, \apj, 659, 1106

\bibitem[{{Muzzin} {et~al}\mbox{.}(2007{\natexlab{b}}){Muzzin}, {Yee}, {Hall},
  \& {Lin}}]{Muzzin-2007b}
{Muzzin} A., {Yee} H.~K.~C., {Hall} P.~B., {Lin} H., 2007{\natexlab{b}}, \apj,
  663, 150

\bibitem[{{Nagai} \& {Kravtsov}(2005)}]{Nagai-2005}
{Nagai} D., {Kravtsov} A.~V., 2005, \apj, 618, 557

\bibitem[{{Navarro}, {Frenk} \& {White}(1997){Navarro}, {Frenk}, \&
  {White}}]{Navarro-1997}
{Navarro} J.~F., {Frenk} C.~S., {White} S.~D.~M., 1997, \apj, 490, 493

\bibitem[{{Navarro} {et~al}\mbox{.}(2004){Navarro}, {Hayashi}, {Power},
  {Jenkins}, {Frenk}, {White}, {Springel}, {Stadel}, \& {Quinn}}]{Navarro-2004}
{Navarro} J.~F. {et~al.}, 2004, \mnras, 349, 1039

\bibitem[{{Neto} {et~al}\mbox{.}(2007){Neto}, {Gao}, {Bett}, {Cole}, {Navarro},
  {Frenk}, {White}, {Springel}, \& {Jenkins}}]{Neto-2007}
{Neto} A.~F. {et~al.}, 2007, \mnras, 381, 1450

\bibitem[{{Peacock} \& {Smith}(2000)}]{Peacock-2000}
{Peacock} J.~A., {Smith} R.~E., 2000, \mnras, 318, 1144

\bibitem[{{Percival} {et~al}\mbox{.}(2001){Percival} {et~al.}}]{Percival-2001}
{Percival} W.~J., {et~al.}, 2001, \mnras, 327, 1297

\bibitem[{{Phleps} {et~al}\mbox{.}(2006){Phleps}, {Peacock}, {Meisenheimer}, \&
  {Wolf}}]{Phleps-2006}
{Phleps} S., {Peacock} J.~A., {Meisenheimer} K., {Wolf} C., 2006, \aap, 457,
  145

\bibitem[{{Poggianti}(1997)}]{Poggianti-1997}
{Poggianti} B.~M., 1997, \aaps, 122, 399

\bibitem[{{Poggianti} {et~al}\mbox{.}(2010){Poggianti}, {De Lucia}, {Varela},
  {Aragon-Salamanca}, {Finn}, {Desai}, {von der Linden}, \&
  {White}}]{Poggianti-2010}
{Poggianti} B.~M., {De Lucia} G., {Varela} J., {Aragon-Salamanca} A., {Finn}
  R., {Desai} V., {von der Linden} A., {White} S.~D.~M., 2010, \mnras, 405, 995

\bibitem[{{Popesso} {et~al}\mbox{.}(2007){Popesso}, {Biviano}, {B{\"o}hringer},
  \& {Romaniello}}]{Popesso-2007}
{Popesso} P., {Biviano} A., {B{\"o}hringer} H., {Romaniello} M., 2007, \aap,
  464, 451

\bibitem[{{Pozzetti} {et~al}\mbox{.}(2010){Pozzetti} {et~al.}}]{Pozzetti-2010}
{Pozzetti} L., {et~al.}, 2010, \aap, 523, A13+

\bibitem[{{Prada} {et~al}\mbox{.}(2006){Prada}, {Klypin}, {Simonneau},
  {Betancort-Rijo}, {Patiri}, {Gottl{\"o}ber}, \& {Sanchez-Conde}}]{Prada-2006}
{Prada} F., {Klypin} A.~A., {Simonneau} E., {Betancort-Rijo} J., {Patiri} S.,
  {Gottl{\"o}ber} S., {Sanchez-Conde} M.~A., 2006, \apj, 645, 1001

\bibitem[{{Rines} \& {Diaferio}(2006)}]{Rines-2006}
{Rines} K., {Diaferio} A., 2006, \aj, 132, 1275

\bibitem[{{Rosati} {et~al}\mbox{.}(2009){Rosati} {et~al.}}]{Rosati-2009}
{Rosati} P., {et~al.}, 2009, \aap, 508, 583

\bibitem[{{Rozo} {et~al}\mbox{.}(2007){Rozo}, {Wechsler}, {Koester}, {McKay},
  {Evrard}, {Johnston}, {Sheldon}, {Annis}, \& {Frieman}}]{Rozo-2007}
{Rozo} E. {et~al.}, 2007, ArXiv Astrophysics e-prints

\bibitem[{{Schechter}(1976)}]{Schechter-1976}
{Schechter} P., 1976, \apj, 203, 297

\bibitem[{{Skrutskie} {et~al}\mbox{.}(2006){Skrutskie}
  {et~al.}}]{Skrutskie-2006}
{Skrutskie} M.~F., {et~al.}, 2006, \aj, 131, 1163

\bibitem[{{Spergel} {et~al}\mbox{.}(2003){Spergel}, {Verde}, {Peiris},
  {Komatsu}, {Nolta}, {Bennett}, {Halpern}, {Hinshaw}, {Jarosik}, {Kogut},
  {Limon}, {Meyer}, {Page}, {Tucker}, {Weiland}, {Wollack}, \&
  {Wright}}]{Spergel-2003}
{Spergel} D.~N. {et~al.}, 2003, \apjs, 148, 175

\bibitem[{{Springel} {et~al}\mbox{.}(2005){Springel}, {White}, {Jenkins},
  {Frenk}, {Yoshida}, {Gao}, {Navarro}, {Thacker}, {Croton}, {Helly},
  {Peacock}, {Cole}, {Thomas}, {Couchman}, {Evrard}, {Colberg}, \&
  {Pearce}}]{Springel-2005}
{Springel} V. {et~al.}, 2005, \nat, 435, 629

\bibitem[{{Stott} {et~al}\mbox{.}(2010){Stott}, {Collins}, {Sahl{\'e}n},
  {Hilton}, {Lloyd-Davies}, {Capozzi}, {Hosmer}, {Liddle}, {Mehrtens},
  {Miller}, {Romer}, {Stanford}, {Viana}, {Davidson}, {Hoyle}, {Kay}, \&
  {Nichol}}]{Stott-Collins-2010}
{Stott} J.~P. {et~al.}, 2010, \apj, 718, 23

\bibitem[{{Stott} {et~al}\mbox{.}(2007){Stott}, {Smail}, {Edge}, {Ebeling},
  {Smith}, {Kneib}, \& {Pimbblet}}]{Stott-2007}
{Stott} J.~P., {Smail} I., {Edge} A.~C., {Ebeling} H., {Smith} G.~P., {Kneib}
  J.-P., {Pimbblet} K.~A., 2007, \apj, 661, 95

\bibitem[{{Strazzullo} {et~al}\mbox{.}(2006){Strazzullo}, {Rosati}, {Stanford},
  {Lidman}, {Nonino}, {Demarco}, {Eisenhardt}, {Ettori}, {Mainieri}, \&
  {Toft}}]{Strazzullo-2006}
{Strazzullo} V. {et~al.}, 2006, \aap, 450, 909

\bibitem[{{Temporin} {et~al}\mbox{.}(2008){Temporin} {et~al.}}]{Temporin-2008}
{Temporin} S., {et~al.}, 2008, \aap, 482, 81

\bibitem[{{Thomas} {et~al}\mbox{.}(2005){Thomas}, {Maraston}, {Bender}, \&
  {Mendes de Oliveira}}]{Thomas-2005}
{Thomas} D., {Maraston} C., {Bender} R., {Mendes de Oliveira} C., 2005, \apj,
  621, 673

\bibitem[{{van der Marel} {et~al}\mbox{.}(2000){van der Marel}, {Magorrian},
  {Carlberg}, {Yee}, \& {Ellingson}}]{van-der-Marel-2000}
{van der Marel} R.~P., {Magorrian} J., {Carlberg} R.~G., {Yee} H.~K.~C.,
  {Ellingson} E., 2000, \aj, 119, 2038

\bibitem[{{Vikhlinin} {et~al}\mbox{.}(2009){Vikhlinin}, {Kravtsov}, {Burenin},
  {Ebeling}, {Forman}, {Hornstrup}, {Jones}, {Murray}, {Nagai}, {Quintana}, \&
  {Voevodkin}}]{Vikhlinin-2009}
{Vikhlinin} A. {et~al.}, 2009, \apj, 692, 1060

\bibitem[{{Whiley} {et~al}\mbox{.}(2008){Whiley}, {Arag{\'o}n-Salamanca}, {De
  Lucia}, {von der Linden}, {Bamford}, {Best}, {Bremer}, {Jablonka}, {Johnson},
  {Milvang-Jensen}, {Noll}, {Poggianti}, {Rudnick}, {Saglia}, {White}, \&
  {Zaritsky}}]{Whiley-2008}
{Whiley} I.~M. {et~al.}, 2008, \mnras, 387, 1253

\bibitem[{{White}, {Hernquist} \& {Springel}(2001){White}, {Hernquist}, \&
  {Springel}}]{White-2001}
{White} M., {Hernquist} L., {Springel} V., 2001, \apjl, 550, L129

\bibitem[{{Yang}, {Mo} \& {van den Bosch}(2008){Yang}, {Mo}, \& {van den
  Bosch}}]{Yang-2008}
{Yang} X., {Mo} H.~J., {van den Bosch} F.~C., 2008, \apj, 676, 248

\bibitem[{{Zehavi}(2004)}]{Zehavi-2004}
{Zehavi}, I. e.~a., 2004, \apj, 608, 16

\bibitem[{{Zheng}, {Coil} \& {Zehavi}(2007){Zheng}, {Coil}, \&
  {Zehavi}}]{Zheng-Coil-2007}
{Zheng} Z., {Coil} A.~L., {Zehavi} I., 2007, \apj, 667, 760

\bibitem[{{Zheng} \& {Weinberg}(2007)}]{Zheng-Weinberg-2007}
{Zheng} Z., {Weinberg} D.~H., 2007, \apj, 659, 1

\end{thebibliography}

\end{document}